\documentclass[trackchanges, twocolumn]{aastex7}

\usepackage{float}
\usepackage{hyperref}
\usepackage{amsmath}
\usepackage{subcaption}
\usepackage{url}

\begin{document}

\title{Possibility of Month-scale Quasi-periodic Oscillations in the Gamma-ray Light Curve of OP 313}

 \author[orcid=0000-0003-2445-9935,gname=Sandeep Kumar, sname='Mondal']{Sandeep Kumar Mondal} 
 \affiliation{Tsung-Dao Lee Institute, Shanghai Jiao Tong University, 1 Lisuo Road, Shanghai, 201210, People’s Republic of China}

 \email[show]{sandeep@sjtu.edu.cn} 

\author[orcid=0000-0001-8716-9412,gname=Shubham, sname='Kishore']{Shubham Kishore} 
\affiliation{Indian Institute of Astrophysics (IIA), 2nd Block, Koramangala, Bengaluru - 560034, India}
\email[show]{amp700151@gmail.com}

\author[orcid=0000-0002-9331-4388,gname=Alok C., sname='Gupta']{Alok C. Gupta} 
 \affiliation{Aryabhatta Research Institute of Observational Sciences (ARIES), Manora Peak, Nainital – 263001, India}
\email{acgupta30@gmail.com} 

\author[orcid=0000-0001-9745-5738,gname=Gwenael, sname='Giacinti']{Gwenael Giacinti} 
\affiliation{Tsung-Dao Lee Institute, Shanghai Jiao Tong University, 1 Lisuo Road, Shanghai, 201210, People’s Republic of China}
\affiliation{School of Physics and Astronomy, Shanghai Jiao Tong University, Shanghai 200240, P. R. China}
\affiliation{Key Laboratory for Particle Physics, Astrophysics and Cosmology (Ministry of Education) \& Shanghai Key Laboratory for Particle Physics and Cosmology, 800 Dongchuan Road, Shanghai, 200240, P. R. China}
\email{gwenael.giacinti@sjtu.edu.cn}

\begin{abstract}
\noindent In this work, we report evidence suggesting the potential future detection of a month-scale quasi-periodic oscillation (QPO) in the gamma-ray light curve of
 OP 313. We analysed almost 16.8 years of \textit{Fermi}-LAT gamma-ray data and applied the Bayesian block method to the monthly-binned light curve. We identified four high-flux states and investigated the possibility of a QPO in the fourth high-flux state (MJD 59482–60832). Using the Weighted Wavelet Z-transform (WWZ) and Lomb–Scargle Periodogram (LSP) methods, we find tentative evidence for a month-scale QPO; however, its detection significance is limited by the small number of observed cycles. With a sufficiently long data set, the QPO may be detected with higher significance in the future. We further explored possible physical origins of this potential QPO and examined several models. We found that a curved-jet model can explain the observed behaviour.

\end{abstract}


\section{Introduction} 
\setcounter{footnote}{0}  

\noindent
Blazars are a subclass of radio-loud (RL) active galactic nuclei (AGN) powered by supermassive black holes (SMBHs) with masses ranging from $10^6 - 10^{10}$ \(M_\odot\) \citep{Rees_1984}. They emit bipolar relativistic jets perpendicular to the accretion disk (AD) or along the polar direction, with one jet aligned closely ($\leq$10$^\circ$) to the observer’s line of sight \citep{Urry:1995mg}, producing strongly Doppler-boosted, non-thermal emission spanning the entire electromagnetic (EM) spectrum from radio to gamma-rays. Blazars are divided into Flat Spectrum Radio Quasars (FSRQs), which show strong emission lines in their composite Optical/UV spectra, and BL Lacertae objects (BL Lacs), which exhibit weak or the absence of spectral features. Their emission shows high and rapid variability in flux on timescales ranging from minutes \citep{Aharonian:2007ig} to years \citep{Raiteri:2013dha}. The broadband spectral energy distributions (SED) of blazars exhibits a characteristic double-hump profile \citep{1998MNRAS.299..433F}: the low-energy hump (optical to X-ray) is attributed to synchrotron radiation from relativistic leptons in the jet, while the high-energy hump in (GeV/TeV) gamma-rays is attributed to various leptonic and/or hadronic based emission processes \citep{1998A&A...333..452K,2003APh....18..593M,2004NewAR..48..367K,2013ApJ...768...54B}.  \\ 
\\
Periodic and quasi-periodic oscillations (QPOs) are common in X-ray binaries but rare in AGNs \citep{Remillard_2006, Gierlinski2008, GuptaAC2014}. Blazar variability is categorised into three timescales: (i) Intra-Day variability (IDV), (ii) Short-Term variability (STV), and (iii) Long-Term variability (LTV). Gamma-ray QPOs in AGNs indicate stable processes beyond stochastic variability, shedding light on jet dynamics and accretion across black hole mass scales. IDV-QPOs are typically linked to accretion disk fluctuations, while STV and LTV QPOs are more often attributed to jet-related processes \citep[e.g.][and references therein]{lachowicz20094_QPO_PKS_2155_304,ArkadiptaSarkar:2020dpn,2022Natur_BLLac_Source_QPO,Prince2023_QPO_PKS034627}. In some cases, such as OJ 287, year-scale periodicities are associated with binary SMBH systems \citep{Sillanpaa1988,Sillanpaa1996,2008Natur.452..851V}. \\ 
\\
Since the launch of the Fermi Gamma-ray Space Telescope (FGST) in 2008, the number of reported QPOs in AGNs has risen notably, supported by improved temporal coverage from multiwavelength campaigns \citep{AvikDas:2022iee}. Early detections were limited to a few sources \citep{Sillanpaa1988, lachowicz20094_QPO_PKS_2155_304}, but subsequent studies have reported QPO-like features across a broad range of energy bands \citep[e.g][and references therein]{PG1553_113_QPO_2015, King:2013_QPO_J1359+4011, graham2015_QPO_PG_1302_102, PKS0301_243_QPO2017, ACGupta:2018sgb, Zhou:2018_QPO_PKS_2247_131, 2022Natur_BLLac_Source_QPO, ArkadiptaSarkar:2020dpn, AbhradeepRoy:2022eue, Prince2023_QPO_PKS034627, Kishore2022_QPO_S4_0954_658}. Systematic searches have also been carried out to identify QPO candidates in gamma-ray light curves \citep{Ren:2022_QPO_Sources, 2020_Bhatta_Niraj_QPO_Sources}. However, some studies caution that many of these QPO claims may result from red noise rather than genuine periodicity \citep{covino2019gamma}. \\
\\
OP 313, also catalogued as B2 1308+326 and 4FGL J1310.5+3221, is a flat-spectrum radio quasar (FSRQ) located at a redshift of 0.997 \citep{schneider2010sloan_redshift} with RA= 197.619$^\circ$ \& DEC= +32.3455$^\circ$, (J2000; \cite{RA_decl_OP_313}), placing it among the most distant known blazars \citep{Liodakis:2018kmv}. The source was first identified as a prominent radio emitter in the Bologna B2 survey \citep{Colla1970}, and was later classified as a blazar based on its flat radio spectrum, high optical polarization, and multiwavelength variability— features indicative of non-thermal emission from a relativistic jet closely aligned with the observer’s line of sight \citep{Stickel1991}. Over the years, this source has been monitored in major gamma-ray catalogues, including the \textit{Fermi}-LAT 4FGL \citep{AbdollahiFermi-LAT:2019yla}, and has shown variable emission in the radio, optical, and X-ray bands \citep{Mohammed:2025_Previous_Study}. \\
\\
OP 313 entered a flaring phase in late 2023. In December, the CTA’s LST-1 prototype detected very high-energy gamma rays from the source \citep{OteroSantos2024OP313}, making it the most distant blazar observed at such energies. This detection triggered extensive multiwavelength follow-up observations, and by early 2025, OP 313 was found to be in an extremely active state across the spectrum, with \textit{Fermi}-LAT, MAGIC, Swift-XRT, and optical telescopes \citep{INAF2025OP313}. In the investigation into whether the blazar OP 313 is a changing-look blazar, \citet{2025ApJ...978..120P} found that it is actually an intrinsic FSRQ that appears as a BL Lac in high-flux states due to enhanced nonthermal emission. On May 14, coordinated optical monitoring recorded an R-band magnitude of 13.05, close to its historical maximum \citep{ATel_17184}. In a recent paper, multiband optical flux and spectral variability of the blazar OP 313 on IDV and STV timescales were reported during its outburst in 2024-2025 \citep{2025ApJ...990..214D}.
Radio observations revealed quasi-periodic polarisation \citep{2011_OPTCIAL_QPO_OP_313}, while separate analyses suggested a roughly seven-year jet precession, likely driven by binary black hole dynamics \citep{2017A&A_OP_313_Radio_Not_exactly_QPO}. Additionally, VLBI astrometry spanning approximately 40 years detected an eight-year positional wobble, providing further evidence for jet precession or a binary supermassive black hole system \citep{2024_Makarov_VLBI_QPO_OP_313}. \\
\\
Although previous studies have explored QPOs in the radio and polarisation domains, no dedicated search has been conducted in the gamma-ray regime. Given OP 313’s exceptional variability, high luminosity, and extensive long-term monitoring, it stands out as a compelling candidate for investigating high-energy QPOs. We analysed $\sim$16.8 years of \textit{Fermi}-LAT gamma-ray data of OP 313, covering the period from August 4, 2008, to June 6, 2025. Using the Bayesian block method \citep{Bayesian_Block_2013}, we identified four major flaring states, Flare-A, B, C, and D (Fig.~\ref{fig:BB_LC_30D}). Our analysis focused on Flare-D due to evidence of a potential QPO. By examining the Flare-D light curve with futher shorter time bins (10, 7, 5, and 1 day; Fig.~\ref{fig:Multibin__Flare_D}), we tentatively observed, for the first time, an approximately 83-day periodicity in the gamma-ray emission from OP 313. \\
\\
The paper is structured as follows: section~\ref{sec:Fermi_LAT_Data_Analysis} outlines the \textit{Fermi}-LAT data acquisition procedures and analysis methodology; section~\ref{sec:Fermi-LAT_Lightcurve} presents the \textit{Fermi}-LAT gamma-ray light curve and identifies flaring episodes; section~\ref{sec:QPO_Analysis} details the search for QPOs and the corresponding results; section~\ref{sec:Discussion} discusses the physical implications of the findings and summarizes the outcomes.

\section{ \textit{Fermi}-LAT Data Acquisition} \label{sec:Fermi_LAT_Data_Analysis}
\noindent
The Fermi Gamma-ray Space Telescope (FGST), formerly known as Gamma-ray Large Area Space Telescope (GLAST), was launched into near-earth orbit on 11\textsuperscript{th} June 2008. It carries two instruments on board: one is the Large Area Telescope (LAT), and the other is the Gamma-ray Burst Monitor (GBM). The LAT is Fermi's primary instrument, which is usually called by \textit{Fermi}-LAT. \textit{Fermi}-LAT is an imaging, pair-conversion, wide-field-of-view, high-energy gamma-ray telescope that can detect photons of energy 20 MeV to more than 300 GeV with a field of view of 2.7 sr at 1 GeV and above \citep{Atwood_2009}. Due to its field of view, \textit{Fermi}-LAT can observe approximately 20\% of the sky at any given moment. In survey mode, it covers the whole sky in two orbits around the Earth (Fermi’s orbital period is $\sim$96 minutes), which takes about
3 hours. \\
\\
The Pass 8 \textit{Fermi}-LAT gamma-ray data for OP 313 were retrieved from the Fermi Science Support Centre (FSSC) data server \citep{fermi_lat_query}, covering more than 16.8 years (4\textsuperscript{th} August 2008 to 6\textsuperscript{th} June 2025). Data were extracted within a circular region of interest with a radius of 30$^\circ$ centered on OP 313, in the energy range from 100 MeV to 500 GeV. \\

\begin{figure*}[t]
        \centering
        \includegraphics[width=.9\linewidth]{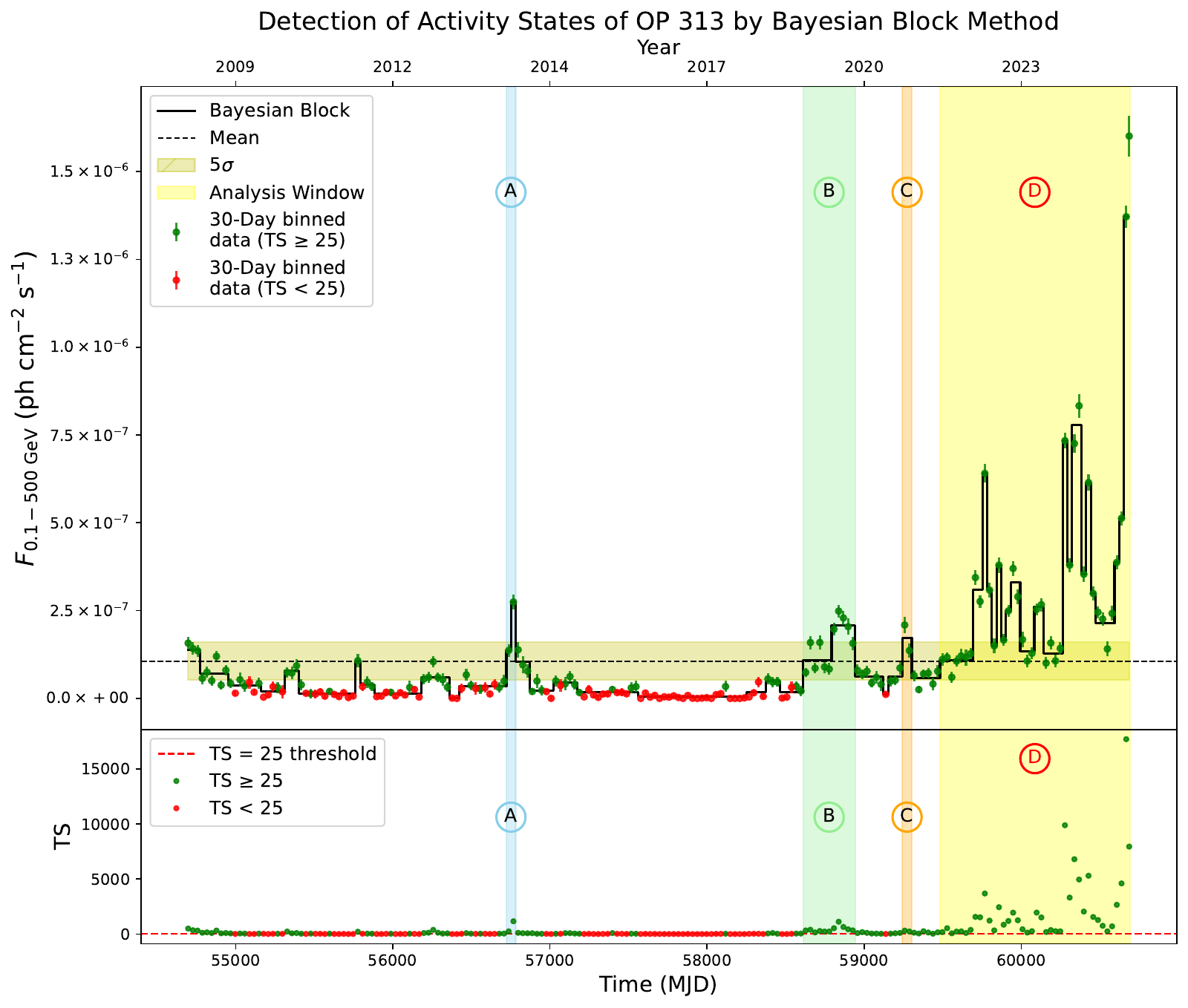}
        \caption{ In the upper plot, four activity states are identified using the Bayesian Block method on a 30-day binned 16.8 years long ( 4\textsuperscript{th} August 2008\- 6\textsuperscript{th} June 2025) \textit{Fermi}-LAT gamma-ray lightcurve of OP313. In the lower plot, the corresponding TS values of the data points have been shown.}
        \label{fig:BB_LC_30D}
    \end{figure*}

\noindent    
The dataset was processed and analysed using Fermipy (v1.0.1; \cite{Fermipy_Version}), an open-source Python package specifically designed for \textit{Fermi}-LAT data analysis. A 10$^\circ\times$10$^\circ$ square analysis region was defined using the `roiwidth' parameter in Fermipy's configuration file, following standard practices outlined in Fermipy’s documentation \footnote{https://fermipy.readthedocs.io/en/latest/config.html}. The analysis was restricted to photon energies between 100 MeV and 500 GeV, and events with zenith angles exceeding 90$^\circ$ were excluded to minimise contamination from Earth limb photons. \\
\\
In the case of event selection, the ‘P8R3 SOURCE’ event class (evclass=128) was chosen, as recommended for analyses involving relatively small region of interest ($<25^\circ$) \citep{bruel2018fermi}. The evtype parameter was set to 3, encompassing all event types (both front and back sections of the tracker). Additionally, data quality cuts were applied using $ \texttt{DATA\_QUAL} > 0 \ \&\& \ \texttt{LAT\_CONFIG} == 1 $
 to ensure inclusion of only high-quality data, acquired under standard LAT science operations. \\
\\
For source modelling, the \textit{Fermi}-LAT Fourth Source Catalogue Data Release 4 (4FGL-DR4; gll\_psc\_v35.fits) \citep{4FGL_DR4} was incorporated. The Galactic diffuse emission was modelled using the latest template gll\_iem\_v07 \citep{Fermi-LAT_galactic_diffuse_model}, while the isotropic extragalactic background was modelled with iso\_p8r3\_source\_v2\_v1.txt. We kept the parameters of the source 4FGL J1310.5+3221, the isotropic diffuse (isodiff) component, and the galactic diffuse (galdiff) component free within the region of interest (ROI). Specifically, the normalisation (norm) parameters were freed for the diffuse components, while for 4FGL J1310.5+3221, modelled by a log-parabola, the normalisation (norm), spectral index (alpha), and curvature (beta) parameters were freed. We then performed the fit using gta.fit() with the NEWMINUIT optimiser, iterating until the fit quality reached 3; the optimiser returned the best-fit model. Subsequently, following the Fermipy user documentation \citep{fermipy_docs}, the \textit{Fermi}-LAT gamma-ray light curve for OP 313 was extracted. \\
\\
The resulting 30-day binned \textit{Fermi}-LAT gamma-ray light curve is shown in the top panel of Fig.~\ref{fig:BB_LC_30D}.

\section{ \textit{Fermi}-LAT Lightcurve} \label{sec:Fermi-LAT_Lightcurve}
\noindent
As mentioned in Section~\ref{sec:Fermi_LAT_Data_Analysis}, we analysed the \textit{Fermi}-LAT gamma-ray light curve spanning 16.8 years, from 4\textsuperscript{th} August 2008 to 6\textsuperscript{th} June 2025 (MJD 54682.65–60832.00). The light curve, shown in the upper panel of Fig.~\ref{fig:BB_LC_30D}, is binned in 30-day intervals. The average/ mean gamma-ray flux over the entire period is indicated by a horizontal black-dashed line in the upper panel. \\
\\
To identify the intervals of enhanced emission, we applied the Bayesian block method \citep{Bayesian_Block_2013} to the light curve (depicted as a black solid line in the upper panel of Fig.~\ref{fig:BB_LC_30D}). A `flare' state is defined as any period during which the gamma-ray flux rises above the mean level. The start and end times of each flare were determined based on the rising and falling segments of the Bayesian block structure along the time axis. This analysis revealed four distinct flaring episodes, labelled Flare-A, Flare-B, Flare-C, and Flare-D, and marked in the figure (Fig.~\ref{fig:BB_LC_30D}) using vertical coloured bars in blue, green, orange, and yellow, respectively. These flares are also identified as regions A, B, C, and D in Fig.~\ref{fig:BB_LC_30D}. This source has recently shown a significantly elevated gamma-ray flux \citep{ATel_16970, ATel_17167}.
Additionally, the detection significance of each time bin is represented using the Test Statistic (TS) values: data points with TS$<$25 are shown in red in Fig.~\ref{fig:BB_LC_30D}, while those with TS$\geq$25 are shown in green. Notably, all data points during the flaring periods have TS$\geq$25, indicating strong detection significance throughout the flare events. \\

\begin{table}
    \centering
        \caption{Average \textit{Fermi}-LAT gamma-ray flux during Flare-A, Flare-B, Flare-C, and Flare-D.}
    \begin{tabular}{||c|c|c||}
    \hline
        Flare & Time  &  Average  \textit{Fermi}-LAT \\
        Name & (MJD)  &  gamma-ray flux  \\
          &   & (ph/$cm^2$/s) \\
         \hline
        Flare-A & 56722- 56782     & 2.06$\times10^{-7}$  \\
        Flare-B & 58612- 58942   & 1.53$\times10^{-7}$   \\
        Flare-C & 59242- 59304  & 1.73$\times10^{-7}$   \\
       Flare-D & 59482- 60832  & 3.42$\times10^{-7}$   \\
       \hline
    \end{tabular}
    \label{tab:Flare_Details}
\end{table}

\begin{figure*}[t]
    \centering
    \includegraphics[width=\linewidth]{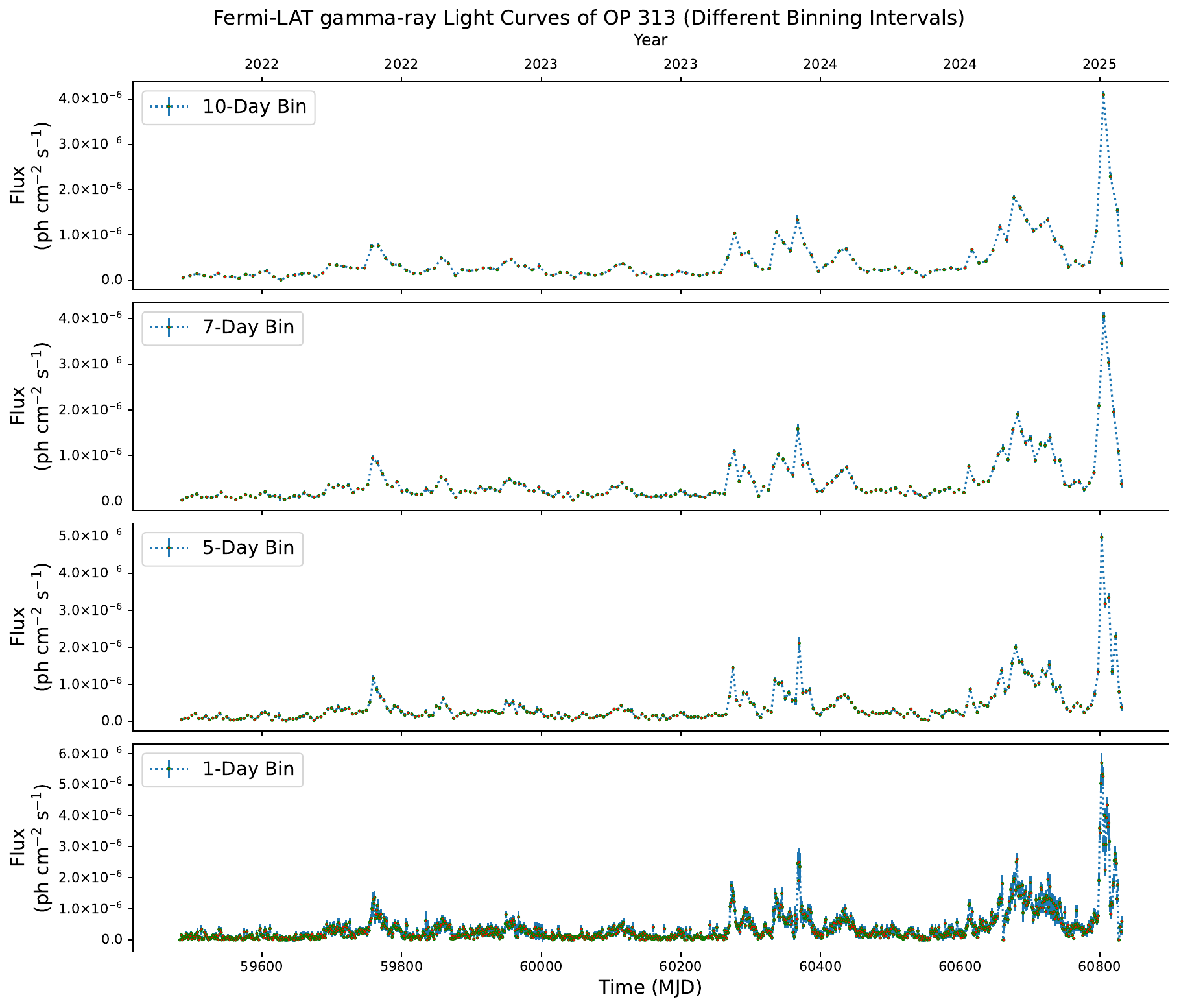}
    \caption{\textit{Fermi}-LAT gamma-ray lightcurve of Flare-D of OP 313 in 10-Day, 7-Day, 5-Day, and 1-Day binning.}
    \label{fig:Multibin__Flare_D}
\end{figure*}

\noindent
In Table~\ref{tab:Flare_Details}, we present the period and average gamma-ray flux of each flare, calculated from 30-day binned \textit{Fermi}-LAT gamma-ray data.
Smaller time bins in the light curve reveal finer sub-structures \citep{Mondal:2021vbs}, which can provide valuable insights for our study. Therefore, we further analysed the \textit{Fermi}-LAT gamma-ray light curve of Flare-D using smaller time bins of 10 days, 7 days, 5 days, and 1 day over the same period, following the same method. The multi-binned light curves are shown in Fig.~\ref{fig:Multibin__Flare_D}. These bin sizes were not chosen arbitrarily; smaller bins reveal more details of the light curve. As can be seen in Fig.~\ref{fig:Multibin__Flare_D}, the structures become more prominent with decreasing bin size. However, flux errors increase as the bin size decreases. Bins smaller than 1 day were not considered because the error bars were already large. We used the 1-day binned \textit{Fermi}-LAT gamma-ray light curve to investigate the presence of QPOs. Details of the QPO analysis and findings are discussed in the following section (section~\ref{sec:QPO_Analysis}).

\section{Analysis and Results} \label{sec:QPO_Analysis}
\subsection{Weighted Wavelet Z (WWZ) analysis}
\begin{figure}
    \centering
    \includegraphics[width=1.05\linewidth]{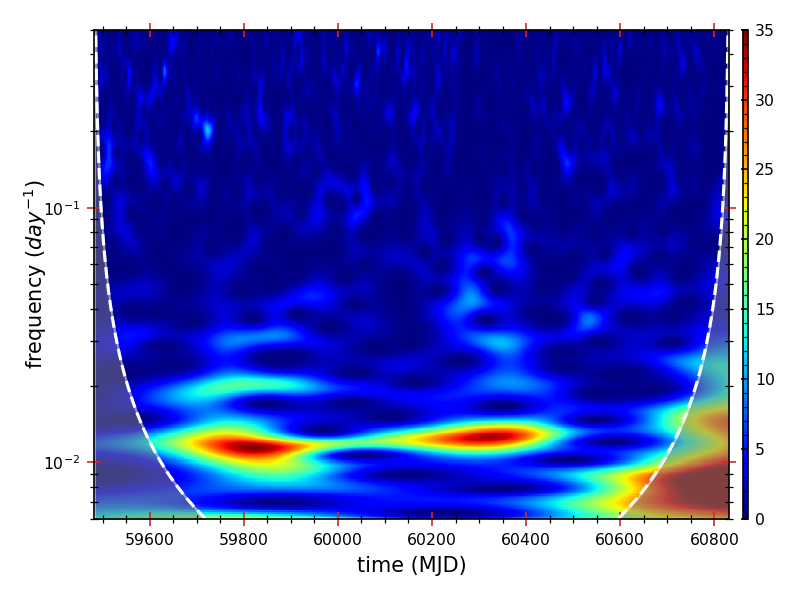}
    \caption{WWZ map for the light curve portion `D' (highlighted in Fig.~\ref{fig:BB_LC_30D})  of OP~313}
    \label{fig:3}
\end{figure}
\begin{figure}
    \centering
    \includegraphics[width=1.0\linewidth]{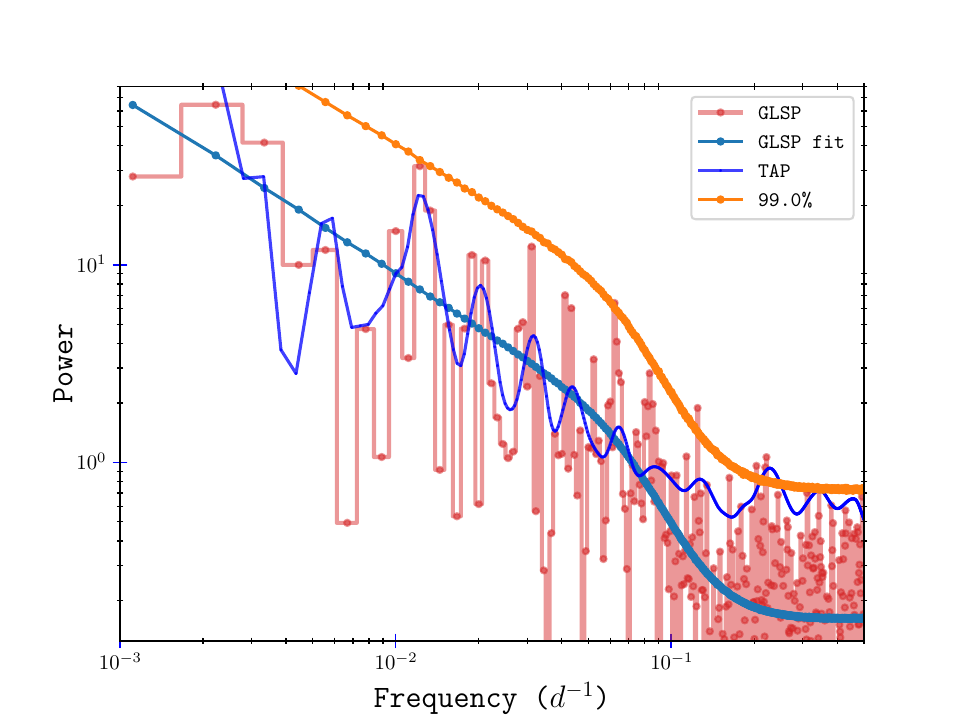}
    \caption{GLSP of the light curve segment MJD~59600 -- MJD~60500}
    \label{fig:4}
\end{figure}
\begin{figure}
    \centering
    \includegraphics[width=1.0\linewidth]{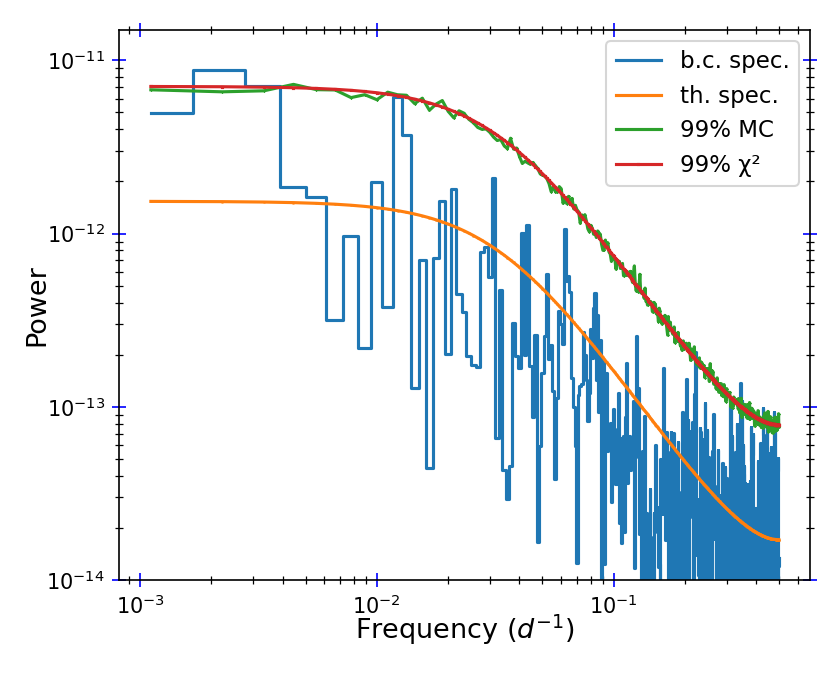}
    \caption{REDFIT analysis of the light curve segment MJD~59600 -- MJD~60500}
    \label{fig:5}
\end{figure}
\noindent
Visual inspection of the 1-day binned light curve of the source reveals notable fluctuations post MJD~59600 and appears to have a roughly periodic interval, suggesting a QPO feature. One of the most robust techniques for inspecting any prominent periodicity features in a time-series is the weighted wavelet-z analysis. This approach has the ability to detect the oscillatory features in the time series along with the time of onset, dislodging, or any evolution of the frequency with time (if the feature is transient). The method relies on the localised wavelets of a mixture of multiple temporal frequencies that dissociates the time series to provide a two-dimensional map of power over frequency and time \citep[or the scale and location, see][and references therein]{1996AJ....112.1709F, 2004JAVSO..32...41T}. The map is computed via convolution of these wavelets and the lightcurve, and is given as \citep[see][]{1996AJ....112.1709F}
\begin{equation}
    W[\tau,\omega;X(t)]=\omega^{1/2}\int X(t)~Y^*(\omega(t-\tau))~dt
\end{equation}
where \(X(t),~ \omega,~\tau~\text{and }Y^*\) represent, respectively, the time series data, test frequency, location of the wavelet, and the complex conjugate of the wavelet function. To compute the map, we have utilised the publicly available Python package {\tt libwwz}\footnote{\url{https://github.com/ISLA-UH/libwwz}}. The 2-D map obtained is helpful in grasping an overview of the time-frequency dependence of the time series, which can be further utilised to constrain the portions in it that visually appear to be peculiar. Another benefit of using this 2-D map is that it can be integrated along the time axis to get an average dependence of the time series on frequency over the time span. Any signature of the periodicity would then be reflected as a peak/ hump in this averaged periodogram (the so-called time-averaged periodogram (TAP)) that gives a prior idea of the peculiar periodicity and that can be tested further via other available statistical approaches.\\
\\ The extreme flexibility of this method lies in the wavelets as any function (or mixture of functions) can be utilized as the `Mother wavelet' to compute the transform and the scales of few particular wavelets can be implicated to the sinusoidal frequency via analytical equations \citep[see][]{1996AJ....112.1709F} frequencies, leading to their explicit utilizations. We have used `Morlet wavelet' (a combination of sinusoidal waveform and a Gaussian window function given as \(Y(z)\sim e^{iz}e^{-cz^2}\); equality is dropped to offer room for additional subdominant factors for proper normalization, if required) as the `Mother wavelet', presenting the obtained 2-D map (for the portion `D' highlighted in Fig.~\ref{fig:BB_LC_30D}) in Fig.~\ref{fig:3} and the TAP from this map in Fig.~\ref{fig:4} for comparison, showing agreement with other independent statistical tools. In Fig.~\ref {fig:3}, we also overplotted a mildly shaded region under white dashed curves, which simply describes the edge effects on different frequencies due to the finiteness of the light curve. This shaded portion is called the region of influence, and though the features detected in such regions are real,  the powers estimated in this region are prone to be affected due to finiteness of the light curve near the edges; hence, and a strong claim for a strong QPO could only be made if any feature resides outside the ROI, having enough number of cycles of oscillation. \\
\\
The WWZ map (Fig.~\ref{fig:3}) of the light curve shows a strongly discerning feature that is persistent from MJD \(\sim\)59600 to up to MJD \(\sim\)60500, contributing to a TAP peak at \(\sim\)0.012~d\(^{-1}\) (in Fig.~\ref{fig:4}). At the epochs in the map when this feature offers maximum powers, it is also associated with some higher frequency features, which, on visual inspection, hint at being probably the higher order harmonics of the main feature.  At the end of epochs, the map displays a huge concentration of powers smeared over a wide range of frequencies, though residing only in the ROI. These features are attributed to the extreme flare/ variability events, as can be seen from the light curve itself.
\subsection{Generalised Lomb Scargle Periodogram (GLSP)}
\noindent
\begin{table*}[]
\caption{Periodogram bending law fitting parameters for ight curve segment MJD 59600 – MJD 60500}
    \centering
    \begin{tabular}{c c c c c c}
        \hline\hline
        Normalization & Bending frequency & lower index & Higher index & Offset\\
        $(A)$ & $(\nu_b)$ & $(\alpha_l)$ & $(\alpha_h)$ & $(c)$\\
        \hline
         $(1.3_{0.1}^{0.2})\times10^{-1}$ & $(7.9_{0.2}^{1.5})\times10^{-2}$ & $(9.35_{0.28}^{0.38})\times10^{-1}$ & $(4.52_{1.06}^{0.38})$ & $(1.61_{0.06}^{0.15})\times10^{-1}$\\
         \hline
    \end{tabular}
    \label{tab:2}
\end{table*}
The Fourier transform is one of the efficient techniques to discern any periodic component in a time-series data. Based on the transform, the generalised Lomb-Scargle periodogram \citep[see][and references therein]{2009A&A...496..577Z} is widely utilised for periodicity searches and power spectral density (PSD or periodogram) analysis of unevenly sampled time series, and is equivalent to fitting a sinusoidal function of the form \(Y = a~cos\ \omega t+b~sin\ \omega t~+~c\).\\
\\
The WWZ map shows a strong feature between the epochs MJD ~59600 and MJD ~60500. Therefore, we segmented the complete light curve between these epochs to compute GLSP, minimizing the contribution from non-periodic components and better estimate the significance of dominant GLSP peak in the periodogram. The GLSP power is given as follows.
\begin{equation}
 P(\omega)=\frac{1}{2\sigma^2}\left[\rm{cosine\, term + sine\, term}\right],
 \end{equation}
where cosine and sine terms are given as
\begin{equation}
\rm{cosine\, term}=\frac{\left[\sum_j (X_j-\bar{X})\rm{cos}\, \omega(t_j-\tau)\right]^2}{\sum_j \rm{cos}^2\omega(t_j-\tau)},
\end{equation}
\begin{equation}
\rm{sine}\, term=\frac{\left[\sum_j (X_j-\bar{X})\rm{sin}\, \omega(t_j-\tau)\right]^2}{\sum_j \rm{sin}^2\omega(t_j-\tau)},
\end{equation}
 and where $t_j$ is the time of a measurement, $X_j$ the corresponding flux value, $\bar{X}$ is the mean of $X_j$, $\omega$ is the frequency, 
and $\tau$ is given through 
\begin{equation}
    \rm{tan}(2\omega\tau)=\frac{\sum_j\rm{sin}\,(2\omega t_j)}{\sum_j\rm{cos}\,(2\omega t_j)}.
\end{equation}
\noindent
The blazar light curve is well known to show a red noise PSD assisted with a flattening nature at the lower and white noise at higher ends \citep[e.g., see ][]{2010MNRAS.402..307V}.
We tested the fitting of the estimated periodogram of the segmented light curve with a simple power law model (\(P(\omega)~=~A\omega^\alpha~+~c\)) and a bending power law model (\(P(\nu)=A\nu^{-a_l}/({1+(\nu/\nu_b)^{a_h-a_l}})+c\)). We found that the inspected light curve preferred the bending power law. Table~\ref{tab:2} includes the obtained bending power law fitting parameters. The corresponding uncertainties on the parameters have been obtained following \citet{2012A&A...544A..80G}.\\
\\
To estimate the significance of the peculiar feature at $\sim0.012~\text{d}^{-1}$, we followed \cite{2013MNRAS.433..907E} to simulate $10^5$ light curve based on the PSD and probability distribution function of the input light curves. The mean of the PSDs of the $10^5$ simulated light curves gives the average spectrum. Fig.~\ref {fig:4} depicts the GLSP (or the PSD) along with the obtained average spectrum and the TAP obtained from the WWZ map. We also plot 99.0 percentile significance level utilizing the $10^5$ PSDs. We found that the peak at $0.012~\text{d}^{-1}$ barely touches the shown 99\% level and hence can't be strongly considered as a QPO feature. The only positive thing (demanding this feature a genuine one) is the agreement of additional high frequency GLSP peaks almost exactly positioned near the corresponding TAP peaks and acting as the higher order harmonics of the one at $0.012~\text{d}^{-1}$.
\subsection{REDFIT analysis}
\noindent
Over the past, multiple attempts have been made to assess the blazar light curves. In this regard, they have often been explained via different versions of multi-order regression methods such as ARMA, ARFIMA, CARFIMA \citep[e.g.,][, and references therin]{2020ApJS..250....1T},  CARMA \citep[e.g.,][]{2018ApJ...863..175G, 2024ApJ...960...11K}, with the simplest one: AR(1), utilised most often. The REDFIT software \citep{2002CG.....28..421S}, a FORTRAN-based package, allows one to assess time series data with the simplest auto-regressive (AR(1)) approach and is efficient in comprehending the red-noise power spectra and to differentiate any genuine periodic feature against multiple significance levels. It also has the ability to account for bias due to uneven sampling in data series. We have utilised this package as another of the complementary tools to check the significance of the feature detected with previous WWZ and LSP methods.\\
\\
The AR(1) approach fits the time series with the function
\begin{equation}Y(t_i) = \sigma_iY(t_{i-1})+\epsilon(t_i)\end{equation}
\begin{equation}\sigma_i = exp(-(t_i-t_{i-1})/\tau)\end{equation}
where \(Y(t_i)\) is the flux at the i$^{th}$ timestamp, \(\sigma~\text{and}~\tau\) are the two AR(1) parameters, \(\epsilon\) is the white noise which has zero mean and variance given as \(1-exp(-2(t_i-t_{i-1})/\tau)\).
Following \citet{}, the power spectrum estimated with the obtained parameters is given as 
\begin{equation}
    G_{rr}(f_j)=G_0\frac{1-\sigma^2}{1-2\sigma~cos(\pi f_j/f_{Nyq}) + \sigma^2}
\end{equation}
where $f_i,~f_{Nyq}~\text{and}~G_0$ are the discrete frequencies, Nyquist frequency, and average spectral amplitude, respectively \citep[see ][for details]{2002CG.....28..421S}. Here, $\sigma$ is the auto-correlation coefficient, given as \(\sigma=exp(-\Delta t/\tau)\) and \(\Delta t=(t_N-t_1)/(N-1)\).\\
\\ The different local significance levels ($= (1-\alpha)\times100\%,~0<\alpha<1$) for the obtained spectrum (based on $\chi^2$ distribution with two degrees of freedom) are then estimated by multiplying the spectra by a factor $-\text{ln}(\alpha)$ \citep[see eq. 15 of ][and related texts for details]{2005A&A...431..391V}. We employed this REDFIT  package and present the results in Fig.~\ref {fig:5}. Another feature of this package is that it also estimates significance levels using percentiles of ensembles using Monte-Carlo simulations. It has also been incorporated in our analysis and has been included in Fig.~\ref {fig:5}. With this method, we also found that the feature under inspection can only marginally touch the 99\% significance level.

\section{Discussion} \label{sec:Discussion}
\noindent The PSD of the selected portion of the blazar light curve follows a bending power law, much differing from the typical simple power law. Also, it shows a high power law index above the bending frequency (in the range of $\sim4$), untypical for blazars, hinting a complexity of the emission process/mechanism. Although the feature (expected to be a QPO) is found to have low significance, we do note that there is a consistent diminution of the maximum flux values at almost regularly perceivable peaks in the light curve. This can substantially decrease the significance of the feature. The regular diminishing is easily explicable as a result of an increase in the viewing angle of the Doppler-boosted emitting region. The presence of higher order harmonics (seen in the WWZ map, TAP, and GLSP) also supports the idea that the observed feature might be a genuine QPO. With the current observation period, the limited number of observed cycles reduces the detection significance of the QPO. A longer monitoring baseline in the future would allow more cycles to be sampled, thereby strengthening the periodic signal if the QPO truly exists. Given the present indications, this appears likely and can be robustly confirmed with future observations.\\
\\
Currently, there are several physical models that can explain the periodic or quasi-periodic behaviour of blazar light curves. In the following part, we have discussed them in brief,
\begin{itemize}
    \item Binary SMBH AGN System: This model can explain year-long QPOs in a binary SMBH AGN system with a total black hole mass of $\sim \mathrm{10^8M}_{\odot}$ and a binary separation on the milli-parsec scale, leading to orbital periods of several years. In such a scenario, the secondary black hole periodically crosses the accretion disk of the primary during its orbit (see Fig. 6 of \cite{Seifine_2024_Binary_SMBH_AGN_QPO} and Fig. 2 of \cite{Valtonen:2025ovs}), perturbing the accretion flow and giving rise to quasi-periodic variability \citep{Valtonen2008}. This model has been explicitly proposed for the blazar OJ 287, which hosts a massive binary SMBH system with an orbital period of $\sim$12 years \citep{Valtonen2008}. Several year-long QPOs in the gamma-ray band have been interpreted within this framework \citep{Fermi-LAT:2015qom, Sandrinelli:2015bmi, Sandrinelli:2015ijk, Sandrinelli:2017gvn, Zhang:2017ear, PKS0301_243_QPO2017, Sandrinelli:2018rdl}.

\item Rotation of the accretion disk hot-spot or spiral shocks or some other non-axisymmetric phenomena around the innermost region of the accretion disk:
The emission from an accretion disk hotspot orbiting close to the innermost stable circular orbit (ISCO) of the SMBH is quasi-thermal. It can account for the observed optical flux modulation. Such optical variability may, in turn, modulate the seed photon field for the external Compton (EC) process in the jet, giving rise to corresponding flux variations in the gamma-ray band \citep{Gupta:2017qyd}. The orbital period of the hotspot is related to the central black hole mass through the following equation
\begin{equation}
\frac{M}{M_\odot}=\frac{3.23\times10^4P}{(r^{3/2}+a)(1+z)} 
\end{equation}
where $P$ is the orbital period in seconds and $z$ is the redshift of the source. For a Schwarzschild black hole ($r=6.0$, $a=0$) and a maximally rotating Kerr black hole ($r=1.2$, $a=0.9982$; \cite{Gupta:2008fc}), the observed $\sim$83-day period implies SMBH masses of $\mathrm{7.2\times10^9 M_\odot}$ and $\mathrm{5.0\times10^{10} M_\odot}$, respectively. No previous study was found that directly measured the black hole mass of OP 313. Usually, the black hole mass of an FSRQ is considered to be within the range of 10$^8$-10$^9$M$_\odot$.
The first value is already very large, and the second exceeds all other SMBH mass estimates. Since the required black hole masses are unreasonably high, it is unlikely that the observed variability feature originates from a hotspot orbiting near the ISCO. 

In addition, this scenario faces two further challenges: (1) from \cite{Bardeen:1972} we can found that for any realistic SMBH mass and spin, the orbital period at the ISCO is typically much shorter ($\sim$hour scale for any black hole of mass of 10$^8$-10$^9$M$_\odot$) than the observed timescale, and (2) the gamma-ray modulation would not exactly match the optical one, since relativistic Doppler boosting alters the observed period in the jet. Another possibility is jet precession, which can generate QPOs in blazar light curves, but the expected timescale is typically longer than a year \citep{Rieger:2004}, inconsistent with the $\sim$83-day QPO.
        
\item \cite{Dong:2020ckw} proposed that QPO might originate from kink instability on a timescale of weeks to months in the jet spine. However, an anti-correlation between the optical polarization degree and the flux was found. Currently, due to the lack of well-sampled optical polarization degree data, this scenario cannot be verified.

\item \cite{Marscher:1992} showed that transient QPOs can also arise from a strong turbulent flow occurring behind a propagating shock or a standing shock. In the post-shock plasma, turbulent eddies form and play a major role in periodically changing the Doppler factor due to their rotation. In our study, the turnover period is $\sim$1126 days (considering the Doppler factor during the flare state of OP 313 to be 27 \citep{Britzen:2017}). This suggests the presence of very large eddies, which is required to explain the observed QPO; on top of that, these structures are random and short-lived to produce the QPO that would not persist for many cycles \citep{Paul_Wiita2011}.

\item OP 313 is an FSRQ-type blazar, whose emission is jet-dominated. So it is highly likely that the QPO is connected to the jet emission. In a binary SMBH system \citep{Valtonen2008, graham2015_QPO_PG_1302_102}, the presence of a secondary SMBH leads to jet precession, which can be attributed to the variation of Lorentz factor along the line of sight of the observer. This is another possible reason for the observed QPO in blazars. According to \cite{Fragile:2008bs}, the Lense-Thirring precession of the disk can influence the jet orientation, which generates QPO of period $\sim$1-2 years \citep{Rieger:2006zc}, which in this case is much higher than the observered QPO of OP 313.

\item Jet-induced quasi-periodicity could also be caused due to the motion of the plasma blob following the internal helical structure of the blazar jet (Fig.~\ref{fig:Curved_Jet_Model_Diagram}). As the blob moves along the helical structure, the viewing angle between the blob and the observer's line of sight changes over time, which causes variations in the Doppler factor \citep{Mohan_Mangalam:2015} and leads to day-to-month-scale quasi-periodicity. 
\par
The gamma-ray QPO observed for OP 313 is found to be within this range. The gamma-ray can be produced within the spherical blob or emission region through inverse-Compton (IC) process; via synchrotron self-Compton (SSC) and/ or external Compton (EC) processes.

        \begin{figure}
            \centering
            \includegraphics[width=0.7\linewidth]{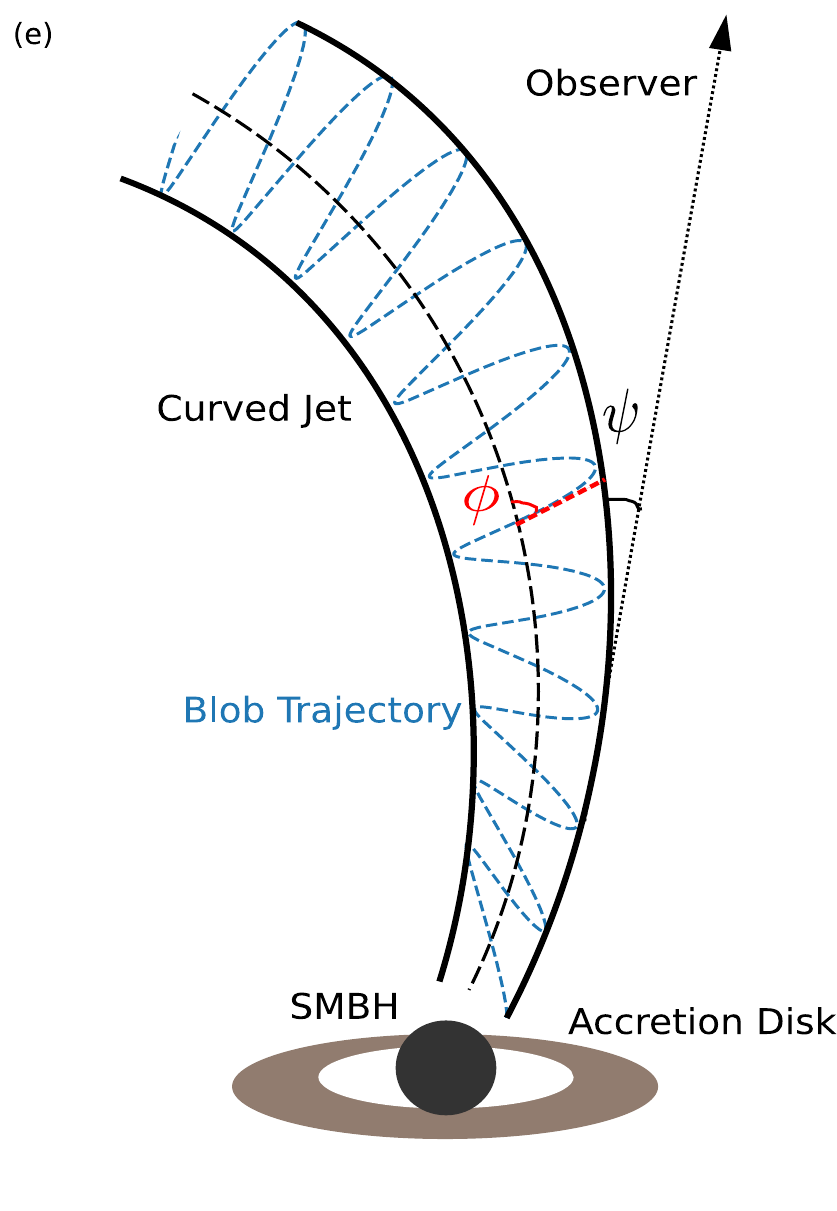}
            \caption{Schematic diagram of blazar curved-jet model (not to scale; \cite{ArkadiptaSarkar:2020dpn}}
            \label{fig:Curved_Jet_Model_Diagram}
        \end{figure}

In Fig.~\ref{fig:Curved_Jet_Model_Diagram}, we have shown a schematic diagram of the curved-jet model, where the spherical blob moves along the helical path (blue dashed line) within the blazar jet (not to scale). The curved jet has launched from the base of the SMBH. The SMBH and accretion disk around the SMBH have also been shown in the diagram. The blob moves outward, i.e., upward in this picture, along the helical path. The angle between the blob velocity vector and the jet axis is $\phi$, also called pitch angle, the angle between the observer's line of sight and the jet axis is $\psi$, and the angle between the blob velocity vector and the observer's line of sight is $\theta$. 
\par
 \begin{figure*}
    \centering

    \begin{subfigure}{0.9\linewidth}
        \centering
        \includegraphics[width=0.9\linewidth]{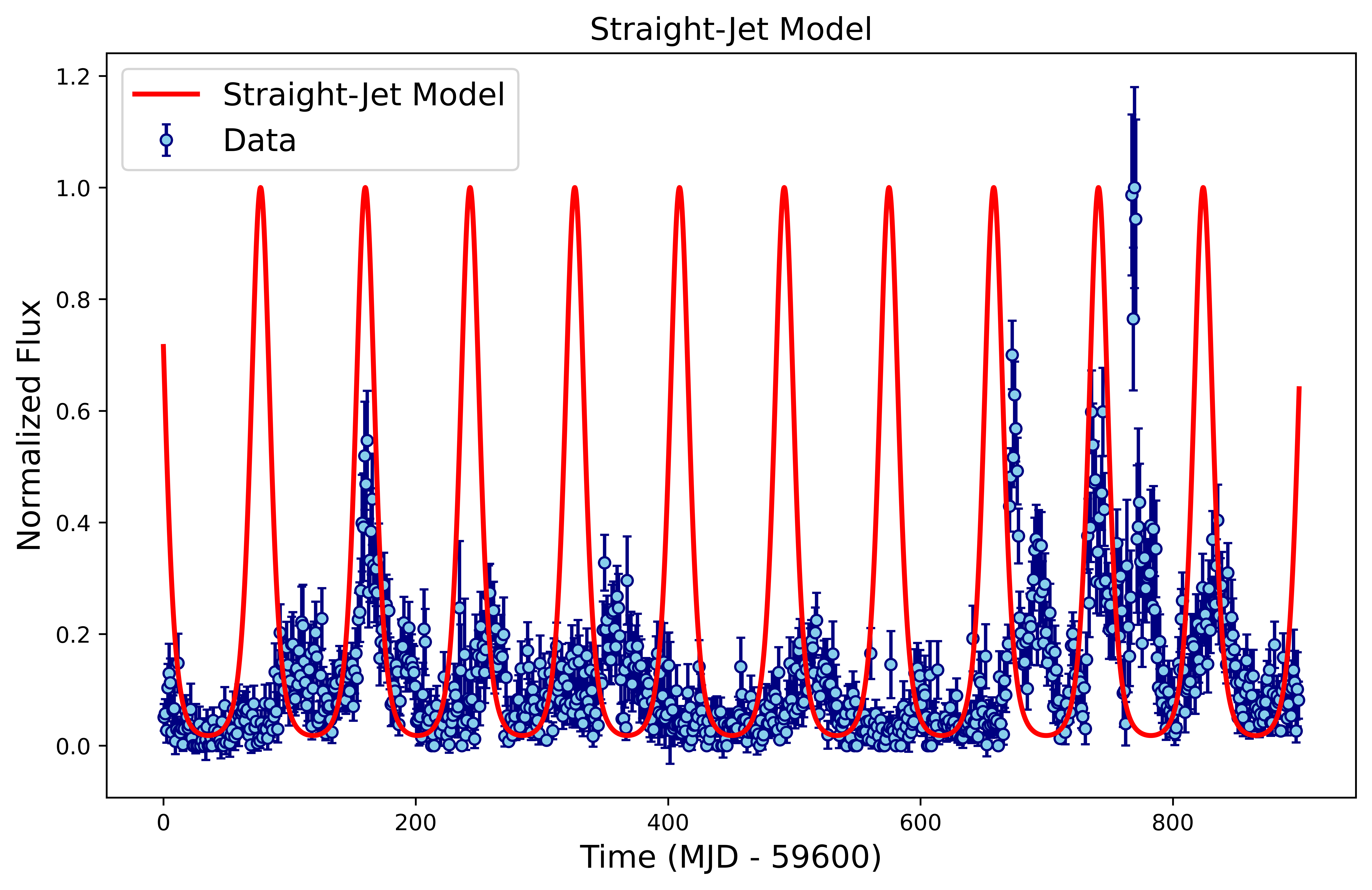}
        \caption{ Simulated light curve using straight-jet model }
        \label{fig:Theoretical_LC_Sim_Straight_Jet}
    \end{subfigure}

    \begin{subfigure}{0.9\linewidth}
        \centering
        \includegraphics[width=0.9\linewidth]{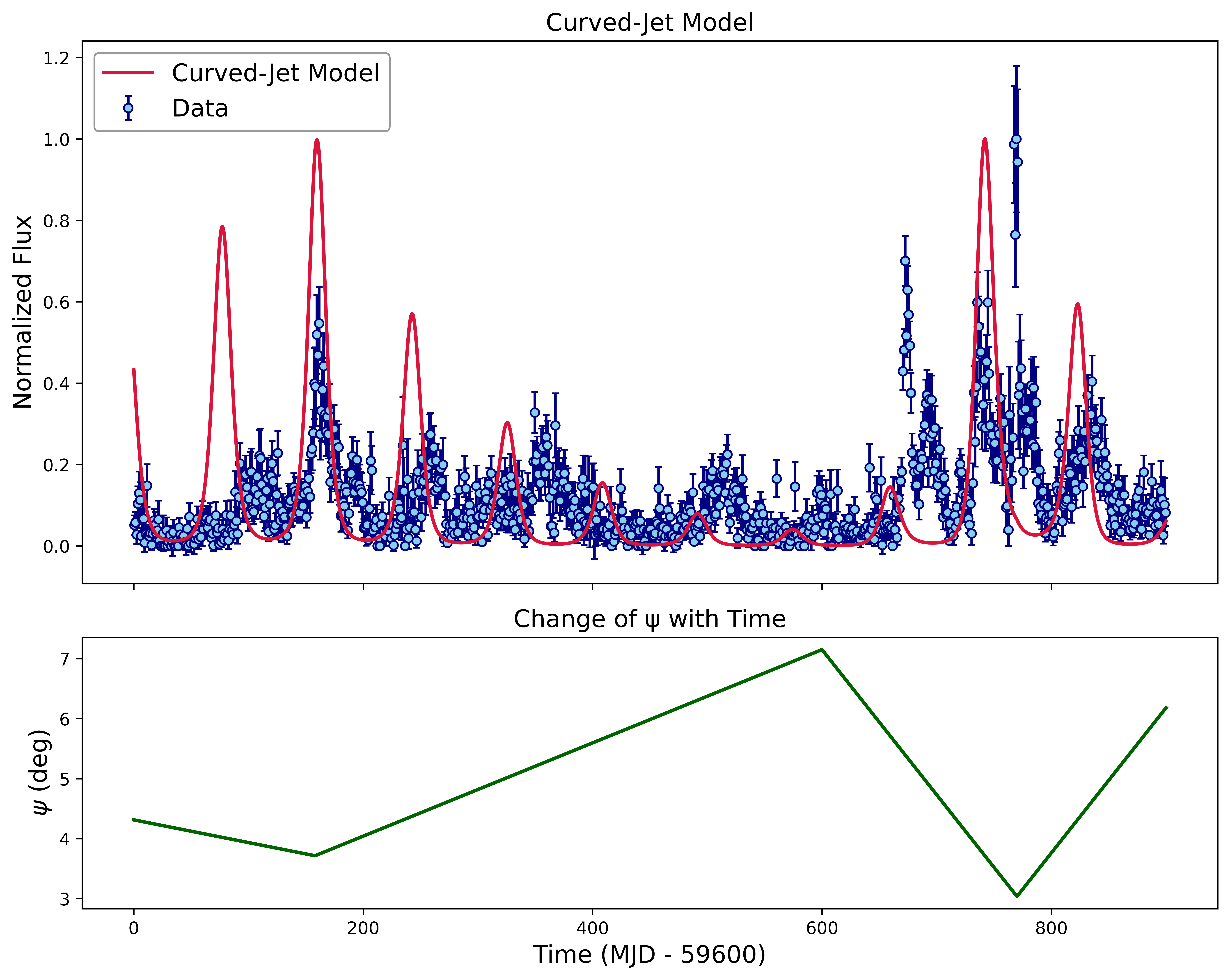}
        \caption{Simulated light curve using curved-jet model and variation of viewing angle of the jet with time}
        \label{fig:Theoretical_LC_Sim_Curved_Jet}
    \end{subfigure}

    \caption{Simulated light curve using straight-jet and curved-jet model}
    \label{fig:Theoretical_LC_Sim}
\end{figure*}

In the case of blazars, the viewing angle of the blob with respect to the observer's line of sight is so small that the observed emission is strongly boosted by relativistic Doppler beaming, which is quantified the the following equation,
\begin{equation} \label{eqn:Doppler_factor}
     \delta=\frac{1}{\Gamma(1-\beta\cos\theta)}
 \end{equation}
 where $\delta$ is the Doppler factor, $\Gamma$ is the bulk Lorentz factor, $\beta$ is the ratio of the bulk velocity of the plasma to the velocity of light in vacuum, and $\theta$ is the angle between the observer's line of sight and the direction of motion of the blob. 
\par
In the case of straight-jet model, the angle between the observer's line of sight and the jet axis remains the same over time, and due to the helical motion of the blob, the viewing angle of the blob with respect to the observer changes, leading to the change in the Doppler factor with time. This causes the periodic fluctuation in the observed gamma-ray flux of blazars (Fig.~\ref{fig:Theoretical_LC_Sim_Straight_Jet}). But, as can be seen in Fig.~\ref{fig:Theoretical_LC_Sim_Straight_Jet}, this model is not able to explain the observed gamma-ray flux variation, so we checked whether the curved-jet model can explain the observed gamma-ray lightcurve or not.
\par
 In the case of the curved-jet model, besides the helical motion of the blob within the blazar jet, the jet is also curved, i.e., the angle between the observer's line of sight and the jet axis changes. The combined effect of these two results not only the periodic fluctuation of the gamma-ray flux but also the amplitude modulation of the observed flux (Fig.~\ref{fig:Theoretical_LC_Sim_Curved_Jet}). Mathematically, $\theta$ changes with time in the following manner, given by \cite{Sobacchi:2017, Zhou:2018_QPO_PKS_2247_131}:
\begin{equation} \label{eqn:cosine_angle}
     \cos\theta(t)=\cos\phi\cos\psi+\sin\psi\sin\phi\cos\Big(\frac{2\pi t}{P_{obs}}\Big)
\end{equation}
 where P$_{obs}$ is the observed period. The observed period can be written as P$_{obs}$ = (1- $\beta \cos\psi \cos\phi$)P$^\prime$, where P$^\prime$ is the period in the blob frame. We set the parameter values of $\phi$ and $\psi$ following the previous studies by \cite{Zhou:2018_QPO_PKS_2247_131,  ArkadiptaSarkar:2020dpn, Prince2023_QPO_PKS034627, Penil:2025pic}. \cite{Britzen:2017} reported the Lorentz factor ($\Gamma$=16.2$\pm$4.4), median Doppler factor ($\bar\delta$=27$\pm$7) of OP 313 using 19 years (1995-2014) of Very Long Baseline Array (VLBA) monitoring data from the MOJAVE survey. \cite{Weaver:2022qty} studied the parsec-scale jet kinematics of a sample of gamma-ray bright blazars monitored over 10 years (2007 June- 2018 December) with the VLBA at 43 GHz under the VLBA-BUBLAZAR program and calculated different physical parameters, e.g., Doppler factor, Lorentz factor. OP 313 or B2 1308+326 is also one of the monitored sources. They quoted the Lorentz factor $\Gamma$= 14.6$\pm$0.7 and Doppler factor $\delta$= 18.5$\pm$2.9 for this source. Also, \cite{Pandey:2023tnt} found that during the low-state the Doppler factor of this source is $\sim$15, which increases to $\sim$27 during the flare state. Our study is focused on the flaring state of this source, so in this work we considered $\delta\sim$27 and $\Gamma$=16.2.  We then estimated $\beta$ using the expression $\Gamma=\frac{1}{\sqrt{1-\beta^2}}$, which was found to be 0.9981, and the QPO period in the blob frame (P$^\prime$) is found to be 42.9 years. Inside the jet, the blob travels 13.09 pc in one cycle. We compute this distance using the formula D=c$\beta$P$^\prime \cos\phi$.


We then simulated the observed flux considering the following mathematical relation,
\begin{equation} \label{eqn:flux_var}
    F_\gamma\propto \delta^\alpha
\end{equation}
where, F$_\gamma$ is the observed gamma-ray flux, $\alpha$ is considered to be 4 \citep{Dermer_Menon:2009} as this approximation is majorly considered while studying the blazar emission. We replaced $\delta$ in Eqn~\ref{eqn:flux_var} by Eqn.~\ref{eqn:Doppler_factor}, and lastly, replaced the cosine function with Eqn.~\ref{eqn:cosine_angle}. In this case, we consider a linear change of the jet viewing angle with respect to the observer, and following previous literature by  \cite{Zhou:2018_QPO_PKS_2247_131,  ArkadiptaSarkar:2020dpn, Prince2023_QPO_PKS034627, Penil:2025pic}, we bounded the range of the angle. Then we modelled the observed lightcurve using this model. 
\par
We used the Akaike information criterion (AIC) to find out the best-fit model statistically, and we found that for the straight-jet model, the AIC value is higher compared to the curved-jet model. Eventually, it can be said that the curved-jet model provides a more likely explanation for the possible gamma-ray quasi-periodicity (Fig.~\ref{fig:Theoretical_LC_Sim_Curved_Jet}).

\end{itemize}


\begin{acknowledgments}
S.K.M. thank A.K. Das for valuable discussions on QPO.
\end{acknowledgments}





%
\facilities{\textit{Fermi}-LAT}

\software{{Fermipy (V1.0.1; \url{https://fermipy.readthedocs.io/en/latest/} ;\citet{Fermipy_Version}), HEAsoft (V6.26.1; \url{https://heasarc.gsfc.nasa.gov/docs/software/heasoft/})
          }}




\bibliography{QPO_OP313}{}

@article{Rees_1984,
   author = "Rees, Martin J.",
   title = "Black Hole Models for Active Galactic Nuclei", 
   journal= "Annual Review of Astronomy and Astrophysics",
   year = "1984",
   volume = "22",
   number = "Volume 22, 1984",
   pages = "471-506",
   doi = "https://doi.org/10.1146/annurev.aa.22.090184.002351",
   url = "https://www.annualreviews.org/content/journals/10.1146/annurev.aa.22.090184.002351",
   publisher = "Annual Reviews",
   issn = "1545-4282",
   type = "Journal Article",
  }

@article{Urry:1995mg,
    author = "Urry, C. Megan and Padovani, Paolo",
    title = "{Unified schemes for radio-loud active galactic nuclei}",
    eprint = "astro-ph/9506063",
    archivePrefix = "arXiv",
    doi = "10.1086/133630",
    journal = "Publ. Astron. Soc. Pac.",
    volume = "107",
    pages = "803",
    year = "1995"
}

@article{Aharonian:2007ig,
    author = "Aharonian, F. and others",
    title = "{An Exceptional Very High Energy Gamma-Ray Flare of PKS 2155-304}",
    eprint = "0706.0797",
    archivePrefix = "arXiv",
    primaryClass = "astro-ph",
    doi = "10.1086/520635",
    journal = "Astrophys. J. Lett.",
    volume = "664",
    pages = "L71--L78",
    year = "2007"
}

@article{Raiteri:2013dha,
    author = "Raiteri, C. M. and Villata, M. and D'Ammando, F. and Larionov, V. M. and Gurwell, M. A. and Mirzaqulov, D. O. and Smith, P. S.",
    collaboration = "GASP-WEBT",
    title = "{The awakening of BL Lacertae: observations by Fermi, Swift, and the GASP-WEBT}",
    eprint = "1309.1282",
    archivePrefix = "arXiv",
    primaryClass = "astro-ph.HE",
    doi = "10.1093/mnras/stt1672",
    journal = "Mon. Not. Roy. Astron. Soc.",
    volume = "436",
    pages = "1530",
    year = "2013"
}

@ARTICLE{Paul_Wiita2011,
       author = {{Wiita}, Paul J.},
        title = "{Quasi-Periodic Oscillations in the X-ray Light Curves of Blazars}",
      journal = {Journal of Astrophysics and Astronomy},
         year = 2011,
        month = jun,
       volume = {32},
       number = {1-2},
        pages = {147-154},
          doi = {10.1007/s12036-011-9071-y},
       adsurl = {https://ui.adsabs.harvard.edu/abs/2011JApA...32..147W}
}

@article{AvikDas:2022iee,
    author = "Das, Avik Kumar and Prince, Raj and Gupta, Alok C. and Kushwaha, Pankaj",
    title = "{The Detection of Possible Transient Quasiperiodic Oscillations in the \ensuremath{\gamma}-Ray Light Curve of PKS 0244-470 and 4C+38.41}",
    eprint = "2211.00588",
    archivePrefix = "arXiv",
    primaryClass = "astro-ph.HE",
    doi = "10.3847/1538-4357/acd17f",
    journal = "Astrophys. J.",
    volume = "950",
    number = "2",
    pages = "173",
    year = "2023"
}

@article{Remillard_2006,
   title={X-Ray Properties of Black-Hole Binaries},
   volume={44},
   ISSN={1545-4282},
   url={http://dx.doi.org/10.1146/annurev.astro.44.051905.092532},
   DOI={10.1146/annurev.astro.44.051905.092532},
   number={1},
   journal={Annual Review of Astronomy and Astrophysics},
   publisher={Annual Reviews},
   author={Remillard, Ronald A. and McClintock, Jeffrey E.},
   year={2006},
   month=sep, pages={49–92} }

@ARTICLE{Gierlinski2008,
       author = {{Gierli{\'n}ski}, Marek and {Middleton}, Matthew and {Ward}, Martin and {Done}, Chris},
        title = "{A periodicity of \raisebox{-0.5ex}\textasciitilde1hour in X-ray emission from the active galaxy RE J1034+396}",
      journal = {\nat},
         year = 2008,
        month = sep,
       volume = {455},
       number = {7211},
        pages = {369-371},
          doi = {10.1038/nature07277},
       adsurl = {https://ui.adsabs.harvard.edu/abs/2008Natur.455..369G}
}

@ARTICLE{GuptaAC2014,
       author = {{Gupta}, Alok C.},
        title = "{Quasi Periodic Oscillations in Blazars}",
      journal = {Journal of Astrophysics and Astronomy},
         year = 2014,
        month = sep,
       volume = {35},
       number = {3},
        pages = {307-314},
          doi = {10.1007/s12036-014-9219-7},
       adsurl = {https://ui.adsabs.harvard.edu/abs/2014JApA...35..307G}
}

@ARTICLE{Sillanpaa1988,
       author = {{Sillanpaa}, A. and {Haarala}, S. and {Valtonen}, M.~J. and {Sundelius}, B. and {Byrd}, G.~G.},
        title = "{OJ 287: Binary Pair of Supermassive Black Holes}",
      journal = {\apj},
         year = 1988,
        month = feb,
       volume = {325},
        pages = {628},
          doi = {10.1086/166033},
       adsurl = {https://ui.adsabs.harvard.edu/abs/1988ApJ...325..628S}
}

@ARTICLE{Sillanpaa1996,
       author = {{Sillanpaa}, A. and {Takalo}, L.~O. and {Pursimo}, T. and {Lehto}, H.~J. and {Nilsson}, K. and {Teerikorpi}, P. and {Heinaemaeki}, P. and {Kidger}, M. and {de Diego}, J.~A. and {Gonzalez-Perez}, J.~N. and {Rodriguez-Espinosa}, J. -M. and {Mahoney}, T. and {Boltwood}, P. and {Dultzin-Hacyan}, D. and {Benitez}, E. and {Turner}, G.~W. and {Robertson}, J.~W. and {Honeycut}, R.~K. and {Efimov}, Yu. S. and {Shakhovskoy}, N. and {Charles}, P.~A. and {Schramm}, K.~J. and {Borgeest}, U. and {Linde}, J.~V. and {Weneit}, W. and {Kuehl}, D. and {Schramm}, T. and {Sadun}, A. and {Grashuis}, R. and {Heidt}, J. and {Wagner}, S. and {Bock}, H. and {Kuemmel}, M. and {Heines}, A. and {Fiorucci}, M. and {Tosti}, G. and {Ghisellini}, G. and {Raiteri}, C.~M. and {Villata}, M. and {de Francesco}, G. and {Bosio}, S. and {Latini}, G.},
        title = "{Confirmation of the 12-year optical outburst cycle in blazar OJ 287.}",
      journal = {\aap},
         year = 1996,
        month = jan,
       volume = {305},
        pages = {L17},
       adsurl = {https://ui.adsabs.harvard.edu/abs/1996A&A...305L..17S}
}

@article{covino2019gamma,
  title={Gamma-ray quasi-periodicities of blazars. A cautious approach},
  author={Covino, Stefano and Sandrinelli, Angela and Treves, Aldo},
  journal={Monthly Notices of the Royal Astronomical Society},
  volume={482},
  number={1},
  pages={1270--1274},
  year={2019},
  publisher={Oxford University Press}
}

@article{lachowicz20094_QPO_PKS_2155_304,
  title={A\~{} 4.6 h quasi-periodic oscillation in the BL Lacertae PKS 2155-304?},
  author={Lachowicz, P and Gupta, AC and Gaur, H and Wiita, PJ},
  journal={Astronomy \& Astrophysics},
  volume={506},
  number={2},
  pages={L17--L20},
  year={2009},
  publisher={EDP Sciences}
}

@article{PG1553_113_QPO_2015,
    author = "Ackermann, M. and others",
    collaboration = "Fermi-LAT",
    title = "{Multiwavelength Evidence for Quasi-periodic Modulation in the Gamma-ray Blazar PG 1553+113}",
    eprint = "1509.02063",
    archivePrefix = "arXiv",
    primaryClass = "astro-ph.HE",
    doi = "10.1088/2041-8205/813/2/L41",
    journal = "Astrophys. J. Lett.",
    volume = "813",
    number = "2",
    pages = "L41",
    year = "2015"
}

@article{King:2013_QPO_J1359+4011,
    author = "King, O. G. and Hovatta, T. and Max-Moerbeck, W. and Meier, D. L. and Pearson, T. J. and Readhead, A. C. S. and Reeves, R. and Richards, J. L. and Shepherd, M. C.",
    title = "{A quasi-periodic oscillation in the blazar J1359+4011}",
    eprint = "1309.1158",
    archivePrefix = "arXiv",
    primaryClass = "astro-ph.HE",
    doi = "10.1093/mnrasl/slt125",
    journal = "Mon. Not. Roy. Astron. Soc.",
    volume = "436",
    pages = "114",
    year = "2013"
}

@article{graham2015_QPO_PG_1302_102,
  title={A possible close supermassive black-hole binary in a quasar with optical periodicity},
  author={Graham, Matthew J and Djorgovski, S George and Stern, Daniel and Glikman, Eilat and Drake, Andrew J and Mahabal, Ashish A and Donalek, Ciro and Larson, Steve and Christensen, Eric},
  journal={Nature},
  volume={518},
  number={7537},
  pages={74--76},
  year={2015},
  publisher={Nature Publishing Group UK London}
}

@article{Zhou:2018_QPO_PKS_2247_131,
    author = "Zhou, Jianeng and Wang, Zhongxiang and Chen, Liang and Wiita, Paul J. and Vadakkumthani, Jithesh and Morrell, Nidia and Zhang, Pengfei and Zhang, Jujia",
    title = "{A 34.5 day quasi-periodic oscillation in gamma-ray emission from the blazar PKS 2247-131}",
    eprint = "1811.02738",
    archivePrefix = "arXiv",
    primaryClass = "astro-ph.HE",
    doi = "10.1038/s41467-018-07103-2",
    journal = "Nature Commun.",
    volume = "9",
    pages = "4599",
    year = "2018"
}

@ARTICLE{2022Natur_BLLac_Source_QPO,
       author = {{Jorstad}, S.~G. and {Marscher}, A.~P. and {Raiteri}, C.~M. and {Villata}, M. and {Weaver}, Z.~R. and {Zhang}, H. and {Dong}, L. and {G{\'o}mez}, J.~L. and {Perel}, M.~V. and {Savchenko}, S.~S. and {Larionov}, V.~M. and {Carosati}, D. and {Chen}, W.~P. and {Kurtanidze}, O.~M. and {Marchini}, A. and {Matsumoto}, K. and {Mortari}, F. and {Aceti}, P. and {Acosta-Pulido}, J.~A. and {Andreeva}, T. and {Apolonio}, G. and {Arena}, C. and {Arkharov}, A. and {Bachev}, R. and {Banfi}, M. and {Bonnoli}, G. and {Borman}, G.~A. and {Bozhilov}, V. and {Carnerero}, M.~I. and {Damljanovic}, G. and {Ehgamberdiev}, S.~A. and {Els{\"a}sser}, D. and {Frasca}, A. and {Gabellini}, D. and {Grishina}, T.~S. and {Gupta}, A.~C. and {Hagen-Thorn}, V.~A. and {Hallum}, M.~K. and {Hart}, M. and {Hasuda}, K. and {Hemrich}, F. and {Hsiao}, H.~Y. and {Ibryamov}, S. and {Irsmambetova}, T.~R. and {Ivanov}, D.~V. and {Joner}, M.~D. and {Kimeridze}, G.~N. and {Klimanov}, S.~A. and {Kn{\"o}tt}, J. and {Kopatskaya}, E.~N. and {Kurtanidze}, S.~O. and {Kurtenkov}, A. and {Kuutma}, T. and {Larionova}, E.~G. and {Leonini}, S. and {Lin}, H.~C. and {Lorey}, C. and {Mannheim}, K. and {Marino}, G. and {Minev}, M. and {Mirzaqulov}, D.~O. and {Morozova}, D.~A. and {Nikiforova}, A.~A. and {Nikolashvili}, M.~G. and {Ovcharov}, E. and {Papini}, R. and {Pursimo}, T. and {Rahimov}, I. and {Reinhart}, D. and {Sakamoto}, T. and {Salvaggio}, F. and {Semkov}, E. and {Shakhovskoy}, D.~N. and {Sigua}, L.~A. and {Steineke}, R. and {Stojanovic}, M. and {Strigachev}, A. and {Troitskaya}, Y.~V. and {Troitskiy}, I.~S. and {Tsai}, A. and {Valcheva}, A. and {Vasilyev}, A.~A. and {Vince}, O. and {Waller}, L. and {Zaharieva}, E. and {Chatterjee}, R.},
        title = "{Rapid quasi-periodic oscillations in the relativistic jet of BL Lacertae}",
      journal = {\nat},
         year = 2022,
        month = sep,
       volume = {609},
       number = {7926},
        pages = {265-268},
          doi = {10.1038/s41586-022-05038-9},
       adsurl = {https://ui.adsabs.harvard.edu/abs/2022Natur.609..265J}
}

@article{PKS0301_243_QPO2017,
    author = "Zhang, Peng-Fei and Yan, Da-Hai and Zhou, Jia-Neng and Fan, Yi-Zhong and Wang, Jian-Cheng and Zhang, Li",
    title = "{A $\gamma$-ray Quasi-Periodic modulation in the Blazar PKS 0301$-$243?}",
    eprint = "1706.02049",
    archivePrefix = "arXiv",
    primaryClass = "astro-ph.HE",
    doi = "10.3847/1538-4357/aa7ecd",
    journal = "Astrophys. J.",
    volume = "845",
    number = "1",
    pages = "82",
    year = "2017"
}

@article{Prince2023_QPO_PKS034627,
    author = "Prince, Raj and Banerjee, Anuvab and Sharma, Ajay and das, Avik Kumar and Gupta, Alok C. and Bose, Debanjan",
    title = "{Quasi-periodic oscillation detected in {\ensuremath{\gamma}}-rays in blazar PKS 0346{\ensuremath{-}}27}",
    eprint = "2308.11317",
    archivePrefix = "arXiv",
    primaryClass = "astro-ph.HE",
    doi = "10.1051/0004-6361/202346400",
    journal = "Astron. Astrophys.",
    volume = "678",
    pages = "A100",
    year = "2023"
}

@article{Kishore2022_QPO_S4_0954_658,
    author = "Kishore, Shubham and Gupta, Alok C. and Wiita, Paul J.",
    title = "{Detection of Quasiperiodic Oscillations in the Blazar S4 0954+658 with TESS}",
    eprint = "2212.08918",
    archivePrefix = "arXiv",
    primaryClass = "astro-ph.HE",
    doi = "10.3847/1538-4357/aca809",
    journal = "Astrophys. J.",
    volume = "943",
    number = "1",
    pages = "53",
    year = "2023"
}

@article{Ren:2022_QPO_Sources,
    author = "Ren, Helena X. and Cerruti, Matteo and Sahakyan, Narek",
    title = "{Quasi-periodic oscillations in the {\ensuremath{\gamma}}-ray light curves of bright active galactic nuclei}",
    eprint = "2204.13051",
    archivePrefix = "arXiv",
    primaryClass = "astro-ph.HE",
    doi = "10.1051/0004-6361/202244754",
    journal = "Astron. Astrophys.",
    volume = "672",
    pages = "A86",
    year = "2023"
}

@ARTICLE{2020_Bhatta_Niraj_QPO_Sources,
       author = {{Bhatta}, Gopal and {Dhital}, Niraj},
        title = "{The Nature of {\ensuremath{\gamma}}-Ray Variability in Blazars}",
      journal = {\apj},
         year = 2020,
        month = mar,
       volume = {891},
       number = {2},
          eid = {120},
        pages = {120},
          doi = {10.3847/1538-4357/ab7455},
archivePrefix = {arXiv},
       eprint = {1911.08198},
 primaryClass = {astro-ph.HE},
       adsurl = {https://ui.adsabs.harvard.edu/abs/2020ApJ...891..120B}
}

@article{schneider2010sloan_redshift,
  title={The sloan digital sky survey quasar catalog. V. Seventh data release},
  author={Schneider, Donald P and Richards, Gordon T and Hall, Patrick B and Strauss, Michael A and Anderson, Scott F and Boroson, Todd A and Ross, Nicholas P and Shen, Yue and Brandt, WN and Fan, Xiaohui and others},
  journal={The Astronomical Journal},
  volume={139},
  number={6},
  pages={2360},
  year={2010},
  publisher={IOP Publishing}
}

@ARTICLE{RA_decl_OP_313,
       author = {{Johnston}, K.~J. and {Fey}, A.~L. and {Zacharias}, N. and {Russell}, J.~L. and {Ma}, C. and {de Vegt}, C. and {Reynolds}, J.~E. and {Jauncey}, D.~L. and {Archinal}, B.~A. and {Carter}, M.~S. and {Corbin}, T.~E. and {Eubanks}, T.~M. and {Florkowski}, D.~R. and {Hall}, D.~M. and {McCarthy}, D.~D. and {McCulloch}, P.~M. and {King}, E.~A. and {Nicolson}, G. and {Shaffer}, D.~B.},
        title = "{A Radio Reference Frame}",
      journal = {\aj},
         year = 1995,
        month = aug,
       volume = {110},
        pages = {880},
          doi = {10.1086/117571},
       adsurl = {https://ui.adsabs.harvard.edu/abs/1995AJ....110..880J}
}

@article{Liodakis:2018kmv,
    author = "Liodakis, I. and Hovatta, T. and Huppenkothen, D. and Kiehlmann, S. and Max-Moerbeck, W. and Readhead, A. C. S.",
    title = "{Constraining the limiting brightness temperature and Doppler factors for the largest sample of radio bright blazars}",
    eprint = "1809.08249",
    archivePrefix = "arXiv",
    primaryClass = "astro-ph.HE",
    doi = "10.3847/1538-4357/aae2b7",
    journal = "Astrophys. J.",
    volume = "866",
    number = "2",
    pages = "137",
    year = "2018"
}

@ARTICLE{Colla1970,
       author = {{Colla}, G. and {Fanti}, C. and {Ficarra}, A. and {Formiggini}, L. and {Gandolfi}, E. and {Grueff}, G. and {Lari}, C. and {Padrielli}, L. and {Roffi}, G. and {Tomasi}, P. and {Vigotti}, M.},
        title = "{A catalogue of 3235 radio sources at 408 MHz}",
      journal = {\aaps},
         year = 1970,
        month = mar,
       volume = {1},
       number = {3},
        pages = {281},
       adsurl = {https://ui.adsabs.harvard.edu/abs/1970A&AS....1..281C}
}

@ARTICLE{Stickel1991,
       author = {{Stickel}, M. and {Padovani}, P. and {Urry}, C.~M. and {Fried}, J.~W. and {Kuehr}, H.},
        title = "{The Complete Sample of 1 Jansky BL Lacertae Objects. I. Summary Properties}",
      journal = {\apj},
         year = 1991,
        month = jun,
       volume = {374},
        pages = {431},
          doi = {10.1086/170133},
       adsurl = {https://ui.adsabs.harvard.edu/abs/1991ApJ...374..431S}
}

@article{AbdollahiFermi-LAT:2019yla,
    author = "Abdollahi, S. and others",
    collaboration = "Fermi-LAT",
    title = "{$Fermi$ Large Area Telescope Fourth Source Catalog}",
    eprint = "1902.10045",
    archivePrefix = "arXiv",
    primaryClass = "astro-ph.HE",
    doi = "10.3847/1538-4365/ab6bcb",
    journal = "Astrophys. J. Suppl.",
    volume = "247",
    number = "1",
    pages = "33",
    year = "2020"
}

@inproceedings{OteroSantos2024OP313,
  author       = {Jorge Otero-Santos and D. Morcuende and M. Nievas Rosillo and D. Sanchez and A. Arbet-Engels and J. Baxter and S. Nozaki and L. Heckmann and E. Visentin and R. De Menezes and F. Di Pierro},
  title        = {The most ancient VHE blazar yet: detection of FSRQ OP 313 at z=0.997 with LST-1},
  booktitle    = {11th International Fermi Symposium},
  year         = {2024},
  month        = sep,
  address      = {College Park, Maryland},
  url          = {https://fermi.gsfc.nasa.gov/science/mtgs/symposia/eleventh/program/tue/Otero-Santos_Fermi_OP313.pdf},
  note         = {Presentation at the 11th International Fermi Symposium},
}

@article{INAF2025OP313,
  author       = {{INAF}},
  title        = {OP 313, the blazar roars again – Media INAF},
  journal      = {INAF Osservatorio Astronomico di Brera},
  year         = {2025},
  month        = feb,
  day          = {9},
  url          = {https://brera.inaf.it/en/inaf-celebrates-its-first-twenty-five-years-media-inaf/},
  note         = {News article on high-energy gamma-ray flare from blazar OP 313},
}

@article{Atwood_2009,
    author = "Atwood, W. B. and others",
    collaboration = "Fermi-LAT",
    title = "{The Large Area Telescope on the Fermi Gamma-ray Space Telescope Mission}",
    eprint = "0902.1089",
    archivePrefix = "arXiv",
    primaryClass = "astro-ph.IM",
    reportNumber = "SLAC-PUB-13620",
    doi = "10.1088/0004-637X/697/2/1071",
    journal = "Astrophys. J.",
    volume = "697",
    pages = "1071--1102",
    year = "2009"
}

@ARTICLE{2018ApJ...863..175G,
       author = {{Goyal}, A. and {Stawarz}, {\L}. and {Zola}, S. and {Marchenko}, V. and {Soida}, M. and {Nilsson}, K. and {Ciprini}, S. and {Baran}, A. and {Ostrowski}, M. and {Wiita}, P.~J. and {Gopal-Krishna} and {Siemiginowska}, A. and {Sobolewska}, M. and {Jorstad}, S. and {Marscher}, A. and {Aller}, M.~F. and {Aller}, H.~D. and {Hovatta}, T. and {Caton}, D.~B. and {Reichart}, D. and {Matsumoto}, K. and {Sadakane}, K. and {Gazeas}, K. and {Kidger}, M. and {Piirola}, V. and {Jermak}, H. and {Alicavus}, F. and {Baliyan}, K.~S. and {Baransky}, A. and {Berdyugin}, A. and {Blay}, P. and {Boumis}, P. and {Boyd}, D. and {Bufan}, Y. and {Campas Torrent}, M. and {Campos}, F. and {Carrillo G{\'o}mez}, J. and {Dalessio}, J. and {Debski}, B. and {Dimitrov}, D. and {Drozdz}, M. and {Er}, H. and {Erdem}, A. and {Escartin P{\'e}rez}, A. and {Fallah Ramazani}, V. and {Filippenko}, A.~V. and {Gafton}, E. and {Garcia}, F. and {Godunova}, V. and {G{\'o}mez Pinilla}, F. and {Gopinathan}, M. and {Haislip}, J.~B. and {Haque}, S. and {Harmanen}, J. and {Hudec}, R. and {Hurst}, G. and {Ivarsen}, K.~M. and {Joshi}, A. and {Kagitani}, M. and {Karaman}, N. and {Karjalainen}, R. and {Kaur}, N. and {Kozie{\l}-Wierzbowska}, D. and {Kuligowska}, E. and {Kundera}, T. and {Kurowski}, S. and {Kvammen}, A. and {LaCluyze}, A.~P. and {Lee}, B.~C. and {Liakos}, A. and {Lozano de Haro}, J. and {Moore}, J.~P. and {Mugrauer}, M. and {Naves Nogues}, R. and {Neely}, A.~W. and {Ogloza}, W. and {Okano}, S. and {Pajdosz}, U. and {Pandey}, J.~C. and {Perri}, M. and {Poyner}, G. and {Provencal}, J. and {Pursimo}, T. and {Raj}, A. and {Rajkumar}, B. and {Reinthal}, R. and {Reynolds}, T. and {Saario}, J. and {Sadegi}, S. and {Sakanoi}, T. and {Salto Gonz{\'a}lez}, J.~L. and {Sameer} and {Simon}, A.~O. and {Siwak}, M. and {Schweyer}, T. and {Sold{\'a}n Alfaro}, F.~C. and {Sonbas}, E. and {Strobl}, J. and {Takalo}, L.~O. and {Tremosa Espasa}, L. and {Valdes}, J.~R. and {Vasylenko}, V.~V. and {Verrecchia}, F. and {Webb}, J.~R. and {Yoneda}, M. and {Zejmo}, M. and {Zheng}, W. and {Zielinski}, P. and {Janik}, J. and {Chavushyan}, V. and {Mohammed}, I. and {Cheung}, C.~C. and {Giroletti}, M.},
        title = "{Stochastic Modeling of Multiwavelength Variability of the Classical BL Lac Object OJ 287 on Timescales Ranging from Decades to Hours}",
      journal = {\apj},
         year = 2018,
        month = aug,
       volume = {863},
       number = {2},
          eid = {175},
        pages = {175},
          doi = {10.3847/1538-4357/aad2de},
archivePrefix = {arXiv},
       eprint = {1709.04457},
 primaryClass = {astro-ph.HE},
       adsurl = {https://ui.adsabs.harvard.edu/abs/2018ApJ...863..175G}
}

@article{4FGL_DR4,
     title={Fermi Large Area Telescope Fourth Source Catalog Data Release 4 (4FGL-DR4)}, 
      author={J. Ballet and P. Bruel and T. H. Burnett and B. Lott and The Fermi-LAT collaboration},
      year={2024},
      journal = {arXiv e-prints},
      eprint={2307.12546},
      archivePrefix={arXiv},
      primaryClass={astro-ph.HE},
      url={https://arxiv.org/abs/2307.12546}, 
}

@ARTICLE{2024ApJ...960...11K,
       author = {{Kishore}, Shubham and {Gupta}, Alok C. and {Wiita}, Paul J.},
        title = "{Rapid Optical Flares in the Blazar OJ 287 on Intraday Timescales with TESS}",
      journal = {\apj},
         year = 2024,
        month = jan,
       volume = {960},
       number = {1},
          eid = {11},
        pages = {11},
          doi = {10.3847/1538-4357/ad0b80},
archivePrefix = {arXiv},
       eprint = {2311.02368},
 primaryClass = {astro-ph.HE},
       adsurl = {https://ui.adsabs.harvard.edu/abs/2024ApJ...960...11K}
}

@ARTICLE{2011_OPTCIAL_QPO_OP_313,
       author = {{Yuan}, Yuhai},
        title = "{Long-term Periodicity Analysis of Polarization Variation for Radio Sources}",
      journal = {Journal of Astrophysics and Astronomy},
         year = 2011,
        month = jun,
       volume = {32},
       number = {1-2},
        pages = {43-46},
          doi = {10.1007/s12036-011-9009-4},
       adsurl = {https://ui.adsabs.harvard.edu/abs/2011JApA...32...43Y}
}

@ARTICLE{2017A&A_OP_313_Radio_Not_exactly_QPO,
       author = {{Qian}, S.~J. and {Britzen}, S. and {Witzel}, A. and {Krichbaum}, T.~P. and {Gan}, H.~Q.},
        title = "{Possible quasi-periodic ejections in quasar B1308+326}",
      journal = {\aap},
         year = 2017,
        month = aug,
       volume = {604},
          eid = {A90},
        pages = {A90},
          doi = {10.1051/0004-6361/201630374},
       adsurl = {https://ui.adsabs.harvard.edu/abs/2017A&A...604A..90Q}
}

@ARTICLE{2024_Makarov_VLBI_QPO_OP_313,
       author = {{Makarov}, Valeri V. and {Lambert}, S{\'e}bastien and {Cigan}, Phil and {DiLullo}, Christopher and {Gordon}, David},
        title = "{Robust 1-norm Periodograms for Analysis of Noisy Non-Gaussian Time Series with Irregular Cadences: Application to VLBI Astrometry of Quasars}",
      journal = {\pasp},
         year = 2024,
        month = may,
       volume = {136},
       number = {5},
          eid = {054503},
        pages = {054503},
          doi = {10.1088/1538-3873/ad4b9f},
archivePrefix = {arXiv},
       eprint = {2405.12324},
 primaryClass = {astro-ph.IM},
       adsurl = {https://ui.adsabs.harvard.edu/abs/2024PASP..136e4503M}
}

@article{ArkadiptaSarkar:2020dpn,
    author = "Sarkar, Arkadipta and Kushwaha, Pankaj and Gupta, Alok C. and Chitnis, Varsha R. and Wiita, Paul J.",
    title = "{Multi-waveband quasi-periodic oscillations in the light curves of blazar CTA 102 during its 2016{\textendash}2017 optical outburst}",
    eprint = "2010.07136",
    archivePrefix = "arXiv",
    primaryClass = "astro-ph.HE",
    doi = "10.1051/0004-6361/202038052",
    journal = "Astron. Astrophys.",
    volume = "642",
    pages = "A129",
    year = "2020"
}

@article{ACGupta:2018sgb,
    author = "Gupta, Alok C. and Tripathi, Ashutosh and Wiita, Paul J. and Kushwaha, Pankaj and Zhang, Zhongli and Bambi, Cosimo",
    title = "{Detection of a quasi-periodic oscillation in gamma-ray light curve of the high redshift blazar B2 1520+31}",
    eprint = "1810.12607",
    archivePrefix = "arXiv",
    primaryClass = "astro-ph.HE",
    doi = "10.1093/mnras/stz395",
    journal = "Mon. Not. Roy. Astron. Soc.",
    volume = "484",
    pages = "5785--5790",
    year = "2019"
}

@article{AbhradeepRoy:2022eue,
    author = "Roy, Abhradeep and others",
    title = "{Detection of a quasi-periodic oscillation in the optical light curve of the remarkable blazar AO 0235+164}",
    eprint = "2205.03586",
    archivePrefix = "arXiv",
    primaryClass = "astro-ph.HE",
    doi = "10.1093/mnras/stac1287",
    journal = "Mon. Not. Roy. Astron. Soc.",
    volume = "513",
    number = "4",
    pages = "5238--5244",
    year = "2022"
}

@article{Mohammed:2025_Previous_Study,
    author = "Mohammed, P. N. Naseef and Aminabi, T. and Baheeja, C. and Sahayanathan, S. and Paliya, Vaidehi S. and Ravikumar, C. D.",
    title = "{Deciphering the multi-wavelength flares of the most distant very high-energy ({\ensuremath{>}}100 GeV) {\ensuremath{\gamma}}-ray emitting blazar}",
    eprint = "2502.01150",
    archivePrefix = "arXiv",
    primaryClass = "astro-ph.HE",
    doi = "10.1016/j.jheap.2025.100365",
    journal = "JHEAp",
    volume = "47",
    pages = "100365",
    year = "2025"
}

@misc{fermi_lat_query,
  author       = {{Fermi Science Support Center}},
  title        = {Fermi LAT Data Server Query},
  howpublished = {\url{https://fermi.gsfc.nasa.gov/cgi-bin/ssc/LAT/LATDataQuery.cgi}},
  note         = {Accessed: 2025-05-03},
  year         = {2025}
}

@misc{fermipy_docs,
  author       = {{Fermipy Collaboration}},
  title        = {Fermipy Documentation (Latest Version)},
  howpublished = {\url{https://fermipy.readthedocs.io/en/latest/}},
  note         = {Accessed: 2025-05-03},
  year         = {2016}
}

@ARTICLE{2020ApJS..250....1T,
       author = {{Tarnopolski}, Mariusz and {{\.Z}ywucka}, Natalia and {Marchenko}, Volodymyr and {Pascual-Granado}, Javier},
        title = "{A Comprehensive Power Spectral Density Analysis of Astronomical Time Series. I. The Fermi-LAT Gamma-Ray Light Curves of Selected Blazars}",
      journal = {\apjs},
         year = 2020,
        month = sep,
       volume = {250},
       number = {1},
          eid = {1},
        pages = {1},
          doi = {10.3847/1538-4365/aba2c7},
archivePrefix = {arXiv},
       eprint = {2006.03991},
 primaryClass = {astro-ph.HE},
       adsurl = {https://ui.adsabs.harvard.edu/abs/2020ApJS..250....1T}
}

@INPROCEEDINGS{Fermipy_Version,
       author = {{Wood}, M. and {Caputo}, R. and {Charles}, E. and {Di Mauro}, M. and {Magill}, J. and {Perkins}, J.~S. and {Fermi-LAT Collaboration}},
        title = "{Fermipy: An open-source Python package for analysis of Fermi-LAT Data}",
    booktitle = {35th International Cosmic Ray Conference (ICRC2017)},
         year = 2017,
       series = {International Cosmic Ray Conference},
       volume = {301},
        month = jul,
          eid = {824},
        pages = {824},
          doi = {10.22323/1.301.0824},
archivePrefix = {arXiv},
       eprint = {1707.09551},
 primaryClass = {astro-ph.IM},
       adsurl = {https://ui.adsabs.harvard.edu/abs/2017ICRC...35..824W}
}

@article{bruel2018fermi,
  title={Fermi-LAT improved Pass\~{} 8 event selection},
  author={Bruel, P and Burnett, TH and Digel, SW and Johannesson, G and Omodei, N and Wood, M},
  journal={arXiv preprint arXiv:1810.11394},
  year={2018}
}

@article{Fermi-LAT_galactic_diffuse_model,
    author = "Acero, F. and others",
    collaboration = "Fermi-LAT",
    title = "{Development of the Model of Galactic Interstellar Emission for Standard Point-Source Analysis of Fermi Large Area Telescope Data}",
    eprint = "1602.07246",
    archivePrefix = "arXiv",
    primaryClass = "astro-ph.HE",
    doi = "10.3847/0067-0049/223/2/26",
    journal = "Astrophys. J. Suppl.",
    volume = "223",
    number = "2",
    pages = "26",
    year = "2016"
}

@ARTICLE{1996AJ....112.1709F,
       author = {{Foster}, Grant},
        title = "{Wavelets for period analysis of unevenly sampled time series}",
      journal = {\aj},
         year = 1996,
        month = oct,
       volume = {112},
        pages = {1709-1729},
          doi = {10.1086/118137},
       adsurl = {https://ui.adsabs.harvard.edu/abs/1996AJ....112.1709F}
}

@ARTICLE{2004JAVSO..32...41T,
       author = {{Templeton}, Matthew},
        title = "{Time-Series Analysis of Variable Star Data}",
      journal = {\jaavso},
         year = 2004,
        month = jun,
       volume = {32},
       number = {1},
        pages = {41-54},
       adsurl = {https://ui.adsabs.harvard.edu/abs/2004JAVSO..32...41T}
}

@ARTICLE{2009A&A...496..577Z,
       author = {{Zechmeister}, M. and {K{\"u}rster}, M.},
        title = "{The generalised Lomb-Scargle periodogram. A new formalism for the floating-mean and Keplerian periodograms}",
      journal = {\aap},
         year = 2009,
        month = mar,
       volume = {496},
       number = {2},
        pages = {577-584},
          doi = {10.1051/0004-6361:200811296},
archivePrefix = {arXiv},
       eprint = {0901.2573},
 primaryClass = {astro-ph.IM},
       adsurl = {https://ui.adsabs.harvard.edu/abs/2009A&A...496..577Z}
}

@ARTICLE{2002CG.....28..421S,
       author = {{Schulz}, Michael and {Mudelsee}, Manfred},
        title = "{REDFIT: estimating red-noise spectra directly from unevenly spaced paleoclimatic time series}",
      journal = {Computers and Geosciences},
         year = 2002,
        month = apr,
       volume = {28},
       number = {3},
        pages = {421-426},
          doi = {10.1016/S0098-3004(01)00044-9},
       adsurl = {https://ui.adsabs.harvard.edu/abs/2002CG.....28..421S}
}

@ARTICLE{2005A&A...431..391V,
       author = {{Vaughan}, S.},
        title = "{A simple test for periodic signals in red noise}",
      journal = {\aap},
         year = 2005,
        month = feb,
       volume = {431},
        pages = {391-403},
          doi = {10.1051/0004-6361:20041453},
archivePrefix = {arXiv},
       eprint = {astro-ph/0412697},
 primaryClass = {astro-ph},
       adsurl = {https://ui.adsabs.harvard.edu/abs/2005A&A...431..391V}
}

@ARTICLE{Bayesian_Block_2013,
       author = {{Scargle}, Jeffrey D. and {Norris}, Jay P. and {Jackson}, Brad and {Chiang}, James},
        title = "{The Bayesian Block Algorithm}",
      journal = {arXiv e-prints},
         year = 2013,
        month = apr,
          eid = {arXiv:1304.2818},
        pages = {arXiv:1304.2818},
          doi = {10.48550/arXiv.1304.2818},
archivePrefix = {arXiv},
       eprint = {1304.2818},
 primaryClass = {astro-ph.IM},
       adsurl = {https://ui.adsabs.harvard.edu/abs/2013arXiv1304.2818S}
}

@ARTICLE{ATel_17184,
       author = {{Marchini}, Alessandro and {Savino}, Joao Pedro Maria and {Stiaccini}, Leonardo and {Verna}, Gaia and {Leonini}, Simone and {Conti}, Massimo and {Rosi}, Paolo and {Ramirez}, Luz Marina Tinjaca and {Bonnoli}, Giacomo and {Peretto}, Ivo and {Lora}, Stefano and {Barbieri}, Matilde and {Rivato}, William and {Sassaro}, Lorenzo and {Agnetti}, Davide},
        title = "{The flaring blazar OP 313 reached an exceptional optical brightness}",
      journal = {The Astronomer's Telegram},
         year = 2025,
        month = may,
       volume = {17184},
        pages = {1},
       adsurl = {https://ui.adsabs.harvard.edu/abs/2025ATel17184....1M}
}

@ARTICLE{ATel_16970,
       author = {{Casaburo}, F. and {Bartolini}, C. and {Ciprini}, S. and {La Mura}, G. and {Monti-Guarnieri}, Pietro},
        title = "{Fermi-LAT detection of renewed gamma-ray activity from the FSRQ sources PKS 0446+11 and OP 313}",
      journal = {The Astronomer's Telegram},
         year = 2025,
        month = jan,
       volume = {16970},
        pages = {1},
       adsurl = {https://ui.adsabs.harvard.edu/abs/2025ATel16970....1C}
}

@ARTICLE{ATel_17167,
       author = {{Zyl}, P. V van and {Monti-Guarnieri}, P.},
        title = "{Fermi-LAT detection of renewed gamma-ray activity from OP 313}",
      journal = {The Astronomer's Telegram},
         year = 2025,
        month = may,
       volume = {17167},
        pages = {1},
       adsurl = {https://ui.adsabs.harvard.edu/abs/2025ATel17167....1Z}
}

@ARTICLE{1998MNRAS.299..433F,
       author = {{Fossati}, G. and {Maraschi}, L. and {Celotti}, A. and {Comastri}, A. and {Ghisellini}, G.},
        title = "{A unifying view of the spectral energy distributions of blazars}",
      journal = {\mnras},
         year = 1998,
        month = sep,
       volume = {299},
       number = {2},
        pages = {433-448},
          doi = {10.1046/j.1365-8711.1998.01828.x},
archivePrefix = {arXiv},
       eprint = {astro-ph/9804103},
 primaryClass = {astro-ph},
       adsurl = {https://ui.adsabs.harvard.edu/abs/1998MNRAS.299..433F}
}

@article{Sandrinelli:2015bmi,
    author = "Sandrinelli, Angela and Covino, Stefano and Dotti, Massimo and Treves, Aldo",
    title = "{Quasi-periodicities at year-like timescales in Blazars}",
    eprint = "1512.04561",
    archivePrefix = "arXiv",
    primaryClass = "astro-ph.GA",
    doi = "10.3847/0004-6256/151/3/54",
    journal = "Astron. J.",
    volume = "151",
    pages = "54",
    year = "2016"
}

@article{Sandrinelli:2015ijk,
    author = "Sandrinelli, Angela and Covino, Stefano and Treves, Aldo",
    title = "{Gamma-ray and optical oscillations in the blazar PKS 0537-441}",
    eprint = "1512.08801",
    archivePrefix = "arXiv",
    primaryClass = "astro-ph.GA",
    doi = "10.3847/0004-637X/820/1/20",
    journal = "Astrophys. J.",
    volume = "820",
    number = "1",
    pages = "20",
    year = "2016"
}

@article{Fermi-LAT:2015qom,
    author = "Ackermann, M. and others",
    collaboration = "Fermi-LAT",
    title = "{Multiwavelength Evidence for Quasi-periodic Modulation in the Gamma-ray Blazar PG 1553+113}",
    eprint = "1509.02063",
    archivePrefix = "arXiv",
    primaryClass = "astro-ph.HE",
    doi = "10.1088/2041-8205/813/2/L41",
    journal = "Astrophys. J. Lett.",
    volume = "813",
    number = "2",
    pages = "L41",
    year = "2015"
}

@article{Sandrinelli:2017gvn,
    author = "Sandrinelli, A. and others",
    title = "{Gamma-ray and optical oscillations of 0716+714, MRK 421, and BL Lacertae}",
    eprint = "1701.04454",
    archivePrefix = "arXiv",
    primaryClass = "astro-ph.HE",
    doi = "10.1051/0004-6361/201630288",
    journal = "Astron. Astrophys.",
    volume = "600",
    pages = "A132",
    year = "2017"
}

@article{Zhang:2017ear,
    author = "Zhang, Pengfei and Yan, Dahai and Liao, Nenghui and Zeng, Wei and Wang, Jiancheng and Cao, Lijia",
    title = "{Possible Quasi-Periodic modulation in the z = 1.1 $\gamma$-ray blazar PKS 0426-380}",
    eprint = "1701.00899",
    archivePrefix = "arXiv",
    primaryClass = "astro-ph.HE",
    doi = "10.3847/1538-4357/aa7465",
    journal = "Astrophys. J.",
    volume = "842",
    number = "1",
    pages = "10",
    year = "2017"
}

@article{Sandrinelli:2018rdl,
    author = "Sandrinelli, Angela and Covino, Stefano and Treves, Aldo and Holgado, A. Miguel and Sesana, Alberto and Lindfors, Elina and Fallah Ramazani, V.",
    title = "{Quasi-periodicities of BL Lacertae objects}",
    eprint = "1801.06435",
    archivePrefix = "arXiv",
    primaryClass = "astro-ph.HE",
    doi = "10.1051/0004-6361/201732550",
    journal = "Astron. Astrophys.",
    volume = "615",
    pages = "A118",
    year = "2018"
}

@ARTICLE{Seifine_2024_Binary_SMBH_AGN_QPO,
       author = {{Seifina}, E.~V.},
        title = "{Active galaxy nuclei: current state of the problem}",
      journal = {Astronomical and Astrophysical Transactions},
         year = 2024,
        month = jan,
       volume = {34},
       number = {3},
        pages = {249-266},
          doi = {10.48550/arXiv.2311.14830},
archivePrefix = {arXiv},
       eprint = {2311.14830},
 primaryClass = {astro-ph.GA},
       adsurl = {https://ui.adsabs.harvard.edu/abs/2024A&AT...34..249S}
}

@article{Gupta:2008fc,
    author = "Gupta, Alok C. and Srivastava, A. K. and Wiita, Paul J.",
    title = "{Periodic Oscillations in the Intra-day Optical Light Curves of the Blazar S5 0716+714}",
    eprint = "0808.3630",
    archivePrefix = "arXiv",
    primaryClass = "astro-ph",
    doi = "10.1088/0004-637X/690/1/216",
    journal = "Astrophys. J.",
    volume = "690",
    pages = "216--223",
    year = "2009"
}

@article{Mohan_Mangalam:2015,
    author = "Mohan, P. and Mangalam, A.",
    title = "{Kinematics of and emission from helically orbiting blobs in a relativistic magnetized jet}",
    eprint = "1503.06551",
    archivePrefix = "arXiv",
    primaryClass = "astro-ph.HE",
    doi = "10.1088/0004-637X/805/2/91",
    journal = "Astrophys. J.",
    volume = "805",
    number = "2",
    pages = "91",
    year = "2015"
}

@article{Valtonen:2025ovs,
    author = "Valtonen, Mauri J. and others",
    title = "{Identifying the Secondary Jet in the RadioAstron Image of OJ 287}",
    eprint = "2510.06744",
    archivePrefix = "arXiv",
    primaryClass = "astro-ph.HE",
    doi = "10.3847/1538-4357/ae057e",
    journal = "Astrophys. J.",
    volume = "992",
    number = "1",
    pages = "110",
    year = "2025"
}

@article{Gupta:2017qyd,
    author = "Gupta, Alok C. and others",
    title = "{A peculiar multiwavelength flare in the blazar 3C 454.3}",
    eprint = "1708.03504",
    archivePrefix = "arXiv",
    primaryClass = "astro-ph.HE",
    doi = "10.1093/mnras/stx2072",
    journal = "Mon. Not. Roy. Astron. Soc.",
    volume = "472",
    number = "1",
    pages = "788--798",
    year = "2017"
}

@ARTICLE{Bardeen:1972,
       author = {{Bardeen}, James M. and {Press}, William H. and {Teukolsky}, Saul A.},
        title = "{Rotating Black Holes: Locally Nonrotating Frames, Energy Extraction, and Scalar Synchrotron Radiation}",
      journal = {\apj},
         year = 1972,
        month = dec,
       volume = {178},
        pages = {347-370},
          doi = {10.1086/151796},
       adsurl = {https://ui.adsabs.harvard.edu/abs/1972ApJ...178..347B}
}

@ARTICLE{Rieger:2004,
       author = {{Rieger}, Frank M.},
        title = "{On the Geometrical Origin of Periodicity in Blazar-type Sources}",
      journal = {\apjl},
         year = 2004,
        month = nov,
       volume = {615},
       number = {1},
        pages = {L5-L8},
          doi = {10.1086/426018},
archivePrefix = {arXiv},
       eprint = {astro-ph/0410188},
 primaryClass = {astro-ph},
       adsurl = {https://ui.adsabs.harvard.edu/abs/2004ApJ...615L...5R}
}

@article{Dong:2020ckw,
    author = "Dong, Lingyi and Zhang, Haocheng and Giannios, Dimitrios",
    title = "{Kink Instabilities In Relativistic Jets Can Drive Quasi-Periodic Radiation Signatures}",
    eprint = "2003.07765",
    archivePrefix = "arXiv",
    primaryClass = "astro-ph.HE",
    doi = "10.1093/mnras/staa773",
    journal = "Mon. Not. Roy. Astron. Soc.",
    volume = "494",
    number = "2",
    pages = "1817--1825",
    year = "2020"
}

@INPROCEEDINGS{Marscher:1992,
       author = {{Marscher}, A.~P. and {Gear}, W.~K. and {Travis}, J.~P.},
        title = "{Variability of Nonthermal Continuum Emission in Blazars}",
    booktitle = {Variability of Blazars},
         year = 1992,
       editor = {{Valtaoja}, E. and {Valtonen}, M.},
        month = jan,
        pages = {85},
       adsurl = {https://ui.adsabs.harvard.edu/abs/1992vob..conf...85M}
}

@article{Pandey:2023tnt,
    author = "Pandey, Ashwani and Kushwaha, Pankaj and Wiita, Paul J. and Prince, Raj and Czerny, Bozena and Stalin, C. S.",
    title = "{Origin of the broadband emission from the transition blazar B2 1308+326}",
    eprint = "2310.05096",
    archivePrefix = "arXiv",
    primaryClass = "astro-ph.HE",
    doi = "10.1051/0004-6361/202347719",
    journal = "Astron. Astrophys.",
    volume = "681",
    pages = "A116",
    year = "2024"
}

@ARTICLE{Valtonen2008,
       author = {{Valtonen}, M.~J. and {Lehto}, H.~J. and {Nilsson}, K. and {Heidt}, J. and {Takalo}, L.~O. and {Sillanp{\"a}{\"a}}, A. and {Villforth}, C. and {Kidger}, M. and {Poyner}, G. and {Pursimo}, T. and {Zola}, S. and {Wu}, J.-H. and {Zhou}, X. and {Sadakane}, K. and {Drozdz}, M. and {Koziel}, D. and {Marchev}, D. and {Ogloza}, W. and {Porowski}, C. and {Siwak}, M. and {Stachowski}, G. and {Winiarski}, M. and {Hentunen}, V.-P. and {Nissinen}, M. and {Liakos}, A. and {Dogru}, S.},
        title = "{A massive binary black-hole system in OJ287 and a test of general relativity}",
      journal = {\nat},
         year = 2008,
        month = apr,
       volume = {452},
       number = {7189},
        pages = {851-853},
          doi = {10.1038/nature06896},
archivePrefix = {arXiv},
       eprint = {0809.1280},
 primaryClass = {astro-ph},
       adsurl = {https://ui.adsabs.harvard.edu/abs/2008Natur.452..851V}
}

@article{Fragile:2008bs,
    author = "Fragile, P. Chris and Meier, David L.",
    title = "{General Relativistic Magnetohydrodynamic Simulations of the Hard State as a Magnetically-Dominated Accretion Flow}",
    eprint = "0810.1082",
    archivePrefix = "arXiv",
    primaryClass = "astro-ph",
    doi = "10.1088/0004-637X/693/1/771",
    journal = "Astrophys. J.",
    volume = "693",
    pages = "771--783",
    year = "2009"
}

@article{Rieger:2006zc,
    author = "Rieger, Frank M.",
    title = "{Supermassive binary black holes among cosmic gamma-ray sources}",
    eprint = "astro-ph/0611224",
    archivePrefix = "arXiv",
    doi = "10.1007/s10509-007-9467-y",
    journal = "Astrophys. Space Sci.",
    volume = "309",
    pages = "271--275",
    year = "2007"
}

@article{Penil:2025pic,
    author = "Penil, P. and Otero-Santos, J. and Banerjee, A. and Buson, S. and Rico, A. and Ajello, M. and Adhikari, S.",
    title = "{Transient quasiperiodic oscillations of Fermi-LAT blazars under the curved jet model}",
    eprint = "2507.03967",
    archivePrefix = "arXiv",
    primaryClass = "astro-ph.HE",
    doi = "10.1051/0004-6361/202555599",
    journal = "Astron. Astrophys.",
    volume = "700",
    pages = "A208",
    year = "2025"
}

@article{Weaver:2022qty,
    author = {Weaver, Zachary R. and Jorstad, Svetlana G. and Marscher, Alan P. and Morozova, Daria A. and Troitsky, Ivan S. and Agudo, Iv{\'a}n and G{\'o}mez, Jos{\'e} L. and L{\"a}hteenm{\"a}ki, Anne and Tammi, Joni and Tornikoski, Merja},
    title = "{Kinematics of Parsec-scale Jets of Gamma-Ray Blazars at 43 GHz during 10 yr of the VLBA-BU-BLAZAR Program}",
    eprint = "2202.12290",
    archivePrefix = "arXiv",
    primaryClass = "astro-ph.HE",
    doi = "10.3847/1538-4365/ac589c",
    journal = "Astrophys. J. Supp.",
    volume = "260",
    number = "1",
    pages = "12",
    year = "2022"
}

@ARTICLE{Britzen:2017,
       author = {{Britzen}, S. and {Qian}, S.-J. and {Steffen}, W. and {Kun}, E. and {Karouzos}, M. and {Gergely}, L. and {Schmidt}, J. and {Aller}, M. and {Aller}, H. and {Krause}, M. and {Fendt}, C. and {B{\"o}ttcher}, M. and {Witzel}, A. and {Eckart}, A. and {Moser}, L.},
        title = "{A swirling jet in the quasar 1308+326}",
      journal = {\aap},
         year = 2017,
        month = jun,
       volume = {602},
          eid = {A29},
        pages = {A29},
          doi = {10.1051/0004-6361/201629999},
       adsurl = {https://ui.adsabs.harvard.edu/abs/2017A&A...602A..29B}
}

@ARTICLE{Sobacchi:2017,
       author = {{Sobacchi}, Emanuele and {Sormani}, Mattia C. and {Stamerra}, Antonio},
        title = "{A model for periodic blazars}",
      journal = {\mnras},
         year = 2017,
        month = feb,
       volume = {465},
       number = {1},
        pages = {161-172},
          doi = {10.1093/mnras/stw2684},
archivePrefix = {arXiv},
       eprint = {1610.04709},
 primaryClass = {astro-ph.HE},
       adsurl = {https://ui.adsabs.harvard.edu/abs/2017MNRAS.465..161S}
}

@BOOK{Dermer_Menon:2009,
       author = {{Dermer}, Charles D. and {Menon}, Govind},
        title = "{High Energy Radiation from Black Holes: Gamma Rays, Cosmic Rays, and Neutrinos}",
         year = 2009,
       adsurl = {https://ui.adsabs.harvard.edu/abs/2009herb.book.....D},
      publisher={Princeton Univerisity Press}
}
\bibliographystyle{aasjournalv7}



\end{document}